\documentclass[twocolumn,natbib]{svjour3}
\usepackage{amsfonts}
\usepackage{natbib} 
\usepackage{graphicx}
 \usepackage{amsmath}
\usepackage{fancyhdr}
\usepackage{enumerate}
\usepackage{subfigure}
\title{Bayesian variable selection for latent class analysis using a collapsed Gibbs sampler}
\author{ Arthur White \and Jason Wyse  \and Thomas Brendan Murphy \thanks{This work is supported by Science Foundation Ireland under the Clique Strategic Research Cluster (SFI/08/SRC/I1407) and the Insight Research Centre (SFI/12/RC/2289).}}
\institute{Arthur White \and Jason Wyse \at
               \\School of Computer Science and Statistics, O'Reilly Institute, Trinity College, Dublin 2, Ireland.\\
              \email{arwhite@tcd.ie, wyseja@tcd.ie}           %  \\
              \and Thomas Brendan Murphy \at
              \\School of Mathematical Sciences, Complex \& Adaptive Systems Laboratory and Insight Research Centre, University College Dublin, Dublin 4, Ireland.
}
\date{ }

\begin{document}
% \titlerunning{Bayesian variable selection for LCA}
 \bibliographystyle{spbasic}

\maketitle

\newcommand{\ctau}{\hbox{\it$\tau$}}
\newcommand{\cov}{\mbox{cov}}
\newcommand{\diag}{\mbox{diag}}
\newcommand{\Var}{\mbox{var}}
\newcommand{\otherwise}{\mbox{otherwise}}
\newcommand{\boldtheta}{\mbox{\boldmath{$\theta$}}}
\newcommand{\boldlambda}{\mbox{\boldmath{$\lambda$}}}
\newcommand{\boldLambda}{\mbox{\boldmath{$\Lambda$}}}
\newcommand{\boldalpha}{\mbox{\boldmath{$\alpha$}}}
\newcommand{\boldbeta}{\mbox{\boldmath{$\beta$}}}
\newcommand{\boldmu}{\mbox{\boldmath{$\mu$}}}
\newcommand{\boldnu}{\mbox{\boldmath{$\nu$}}}
\newcommand{\boldOmega}{\mbox{\boldmath{$\Omega$}}}
\newcommand{\boldSigma}{\mbox{\boldmath{$\Sigma$}}}
\newcommand{\boldrho}{\mbox{\boldmath{$\rho$}}}
\newcommand{\boldomega}{\mbox{\boldmath{$\omega$}}}
\newcommand{\boldtau}{\mbox{\boldmath{$\tau$}}}
\newcommand{\boldphi}{\mbox{\boldmath{$\phi$}}}
\newcommand{\mle}{\mbox{m.l.e.}}
\newcommand{\iid}{\mbox{i.i.d.}}
\newcommand{\Like}{\mathcal{L}}
\newcommand{\diff}{\mathrm{d}}
\newcommand{\normal}{\mbox{N}}
\newcommand{\by}{\mathbf{y}}
\newcommand{\be}{\mathbf{e}}
\newcommand{\bu}{\mathbf{u}}
\newcommand{\indicator}{\mathrm{I}}
\newcommand{\bX}{\mathbf{X}}
\newcommand{\bx}{\mathbf{x}}
\newcommand{\bV}{\mathbf{V}}
\newcommand{\bz}{\mathbf{z}}
 \newcommand{\bZ}{\mathbf{Z}}
\newcommand{\VO}{$\dot{\mbox{V}}\mbox{O}_2 $ \hspace{0.1mm}}
\newcommand{\VCO}{$\dot{\mbox{V}}\mbox{CO}_2$ \hspace{0.1mm}}
\newcommand{\BF}[1]{{\mathcal BF}_{#1, \overline{#1}}}
\begin{abstract}
Latent class analysis is used to perform model based clustering for multivariate categorical responses. Selection of the variables most relevant for clustering is an important task which can affect the quality of clustering considerably. This work considers a Bayesian approach for selecting the number of clusters and the best clustering variables. The main idea is to reformulate the problem of group and variable selection as a probabilistically driven search over a large discrete space using Markov chain Monte Carlo (MCMC) methods. Both selection tasks are carried out simultaneously using an MCMC approach based on a collapsed Gibbs sampling method, whereby several model parameters are integrated from the model, substantially improving computational performance. Post-hoc procedures for parameter and uncertainty estimation are outlined. The approach is tested on simulated and real data.
\end{abstract}

\section{Introduction}\label{sec:intro}
Latent class analysis (LCA) models \citep{goodman1974} are used to discover affinities and groupings in multivariate categorical data. An example of data for which a LCA model may be appropriate would be records for $N$ items on $M$ different variables where each of the variables observed is categorical, more precisely, binary or nominal. There may be a varying number of categories over the $M$ different variables. In using a LCA model one expects the items to segment into groups within which items are similar in nature. These groupings are captured by allowing the variables' multinomial probabilities to vary by group. 

LCA models have been widely studied and applied \citep{aitkin1981,garrett2000,Walsh06}. They can be viewed as a special case of model-based clustering \citep{mclachlan02, fraley07}, in which each item is assumed to arise from one of a number of groups, each group having its own data probability distribution. As with other applications of clustering, two of the main difficulties when formulating a LCA model are identifying a suitable number of different groups or clusters in the data, and choosing the variables which are most informative for groupings in the data.

\citet{dean10} have demonstrated that both of these aspects of model selection can have a considerable effect on the resulting clustering of the data, and that these questions are important in the formulation and application of LCA models. Their approach learns from the data using a headlong search algorithm to choose the optimal clustering variables and number of groups, with steps in the algorithm determined using the Bayesian Information criterion (BIC) \citep{schwarz1978,raftery1995}. 

In this paper, a Bayesian approach for the analysis of LCA models is proposed. It is shown how simple marginalization of the parameters in a LCA model leads to a form of the model for which Markov chain Monte Carlo (MCMC) sampling algorithms can be used to quantify precisely the uncertainty in the number of groups in the data, as well as which variables give the best clustering. This is similar to work carried out by \citet{nobile07} in the analysis of Gaussian finite mixtures, \citet{wyse10} in a block clustering application, \citet{mcdaid2013} in the clustering of social network data, and \citet{tadesse2005} in the context of clustering variable selection for Gaussian distributed data.

The advantage of such an approach is that it allows us to see which of the variables hold information on the clustering of the data. This gives more information than would be available from the headlong search algorithm in \citet{dean10} which gives a point estimate of the variables which optimally determine the clustering, but does not include any quantification of the uncertainty around these particular choices.

The rest of this paper is organised as follows. Section~\ref{sec:lca} outlines model specification for LCA in detail and gives a brief overview of other approaches for analysis that have been proposed in the literature. The five datasets which the method is applied to are then introduced and described.  Section~\ref{sec:collapse} presents the adopted marginalization approach and gives details of the MCMC algorithm used to estimate the models. Section \ref{sec:post-hoc} discusses estimation of posterior quantities from the model, including parameter estimation, how model comparison may be performed by computing approximate posterior model probabilities, as well as a description of how label switching is accounted for. Section~\ref{sec:apply} applies the sampling algorithm to simulated and real data, concluding with a discussion in Section~\ref{sec:discuss}. %Some additional results are included in supplementary material available online.

\section{The classic LCA model}\label{sec:lca}

Denote the data by an $N\times M$ matrix $\bX$ where each row is a record of responses for $M$ categorical variables for one item. Row $n$ of $\bX$ is $\bX_n = (X_{n1},X_{n2},\dots,X_{nM})$. The entry $X_{nm}$ takes a value from one of the categories $\{1,2,\dots,C_m\}$ for variable $m$. The LCA model assumes that the $\bX_n$ arise independently from a finite mixture model with $G$ components or classes,
\begin{equation*}
p(\bX_n | \boldtau, \boldtheta, G ) = \sum_{g=1}^G \tau_g p(\bX_n|\boldtheta_g).
%\bx_n \sim \sum_{g=1}^G \tau_g p(\bx_n|\boldtheta_g).
\end{equation*}
The $\tau_g$ are mixture weights, with $\sum_{g=1}^G \tau_g = 1$ and $\tau_g > 0$, for all $g$. Each component has its own set of parameters $\boldtheta_g$ which embody the differences between groups; this holds the multinomial probabilities for all variables for class $g$.
The parameter $\theta_{gmc}$ corresponds to the probability that an item takes category $c$ for variable $m$ within class $g$. For the component data likelihood, a {\it local independence} assumption is made. This assumes that, conditional on an observation's group membership, all variables are independent of each other, so that
\begin{equation*}
p(\bX_n |\boldtheta_g) = \prod_{m=1}^M \prod_{c=1}^{C_m} \theta_{gmc}^{\indicator(X_{nm} = c)},
\end{equation*}
where  $${\indicator(X_{nm} = c)}  = \left \lbrace  \begin{array}{ll} 1 &\mbox{if $X_{nm} = c$} \\  0 & \mbox{otherwise.}\end{array} \right .$$ 

It is convenient to work with the completed data, that is, data augmented with class labels for each item. Denote these by $\bZ$, where $$ Z_{ng} =  \left \lbrace  \begin{array}{ll} 1 &\mbox{if observation $n$ belongs to group $g$} \\  0 & \mbox{otherwise.}\end{array} \right .$$ The completed data likelihood for an observation may then be written

\begin{equation*}
p(\bX_n, \bZ_n| \boldtheta, \boldtau, G) = \prod_{g=1}^G \left \lbrace \tau_g^{} \prod_{m=1}^M \prod_{c=1}^{C_m} \theta_{gmc}^{\indicator(X_{nm} = c)} \right \rbrace^{Z_{ng}}.
\end{equation*}

\subsection{Approaches for analysis}\label{sec:prev}
When the number of groups and variables is assumed fixed, several techniques for performing LCA are available. When attempting to identify the number of groups in the data, models are fitted over a range of groups, with the best fit often determined with the use of an information criterion. 

In a frequentist paradigm,  an expectation-maximisation (EM) algorithm \citep{dempster77}  can be employed.  The BIC or Akaike's information criterion (AIC) \citep{akaike1973} can then be used to identify an optimal model. Goodness-of-fit statistics such as the likelihood ratio test \citep{goodman1974} can also be employed, but may prove difficult to apply to sparse data with a large number of variables \citep{aitkin1981}.  

In a Bayesian paradigm, for a fixed model, a Gibbs sampling \citep{geman84,garrett2000} technique can be used. When several competing models are possible, additional inference methods must also be used to take account of this additional model uncertainty. \citet{garrett2000} propose graphical tools to aid model selection, and detect weak identifiability. If the difference between these distributions is judged to be small, this then suggests that too large a number of groups has been fitted to the data. \citet[Chapter 5]{Fruhwirth2006} contains an excellent overview of many of the following methods, which take a more principled approach.

Monte Carlo based integration of the marginal likelihood can be used to inform model selection \citep{bensmail97}. \citet{Fruhwirth2004} demonstrates this approach for mixture models using importance sampling~\citep{Geweke1989} or the more general bridge sampling methods~\citep{Meng1996}, while \citet{Chopin2010} use a nested sampling approach. Using a harmonic mean estimator~\citep{Newton1994}, \citet{raftery07}, propose the information criteria AIC or BIC Monte Carlo (AICM or BICM). An alternative information criterion, the deviance information criterion~\citep{Spiegelhalter02}, has also been proposed, although this approach has been criticized for its somewhat opaque specification; this can lead to different results depending on its interpretation when used in a mixture model setting \citep{Celeux2006}. 

%Another approach for Bayesian model selection is to use trans-dimensional MCMC techniques such as reversible jump MCMC (RJMCMC) methods~\citep{Green1995} or Markov birth-death processes~\citep{Stephens2000a,Cappe2003}. RJMCMC have been extended to univariate and multivariate finite Gaussian mixture models \citep{Richardson1997,Dellaportas2006}, but to our knowledge have never been applied to categorical data. \citet{tadesse2005} make use of this method to also perform variable selection, with some parameters integrated from the model. We base our approach on an alternative method, a fully collapsed sampler first proposed by \citet{nobile07}. 

Another approach for Bayesian model selection is to use trans-dimensional MCMC techniques such as reversible jump MCMC (RJMCMC) methods~\citep{Green1995} or Markov birth-death processes~\citep{Stephens2000a,Cappe2003}. While RJMCMC has previously been extended to univariate and multivariate finite Gaussian mixture models\citep{Richardson1997,Dellaportas2006}, only recently has it been applied to latent class analysis \citep{JiaChiun2013,Pandolfi2014}; however, these approaches do not consider variable selection.  \citet{tadesse2005} make use of the RJMCMC method to also perform variable selection, with some parameters integrated from the model. We base our approach on an alternative method, a fully collapsed sampler first proposed by  \citet{nobile07}.

%Trans-dimensional MCMC \citep{Stephens2000a}. Monte Carlo based integration of the marginal likelihood. Harmonic mean estimator \citep{Newton1994,raftery07}. Importance sampling \citep{Geweke1989} or the more general bridge sampling \citep{Meng1996,Fruhwirth2004}. \citet[Chapter 5]{Fruhwirth2006} contains an excellent overview of these methods.

Technical issues can also occur when fitting LCA, such as underestimation of standard errors when using the EM algorithm \citep{bartholomew99,Walsh06}, and label-switching for Gibbs sampling \citep{marin05}. The latter issues, and the methods used to avoid them, are discussed in Section~\ref{sec:post-hoc}.

\subsection{Datasets to be analysed}\label{sec:data_description}
We provide a description of the five datasets which are analysed in Section~\ref{sec:apply}. The first examples are simulated data, while two of the three real datasets have been previously analysed using the LCA methods described in Section~\ref{sec:prev}. The last dataset we examine has not been analysed using LCA before now. We examine both binary and non-binary examples. Four of our examples are binary, so that $C_m =2, \mbox{ and } X_{nm} \in \lbrace 1, 2\rbrace$ for all possible values of $n$ and $m$. One of the examples is non-binary, and has $C_m$ varying from 2 up to 5 over the different variables. %However, the software developed from the outlined method in Section~\ref{sec:collapse} can just as easily be applied to categorical data where $C_m \geq 2.$ 

% several different data values are possible.

\subsubsection*{Dean and Raftery simulated datasets} 
We first test our approach on simulated datasets as specified by \citet{dean10}. The first is a binary two class model of $N=500$ observations with 4 informative variables (1-4) and 9 noise variables (5-13), with class weights set to $\tau_1 = 0.6$ and $\tau_2 = 0.4$. The specified parameter values for $\boldsymbol{\theta}$ are given in Table~\ref{tab:simdata}. 

The second simulated dataset is a non-binary three class model with $N=1000$ observations. Again there are 4 informative variables (1-4) and 6 noise variables (5-10). The class weights are $\tau_1=0.3$,  $\tau_2 = 0.4$ and $\tau_3 =0.3$. The values for $\boldsymbol{\theta}$ for the informative variables are given in Table~\ref{tab:simdata21} and those for the non-informative variables are given in Table~\ref{tab:simdata22}.

\begin{table}[ht]
\centering
\caption{Multinomial probabilities $\boldsymbol{\theta}$ in the Dean and Raftery simulated dataset.}
\begin{tabular}{rrr}
  \hline
Variable & Class 1 & Class 2 \\ 
  \hline
1 & 0.6 & 0.2 \\ 
  2 & 0.8 & 0.5 \\ 
  3 & 0.7 & 0.4 \\ 
  4 & 0.6 & 0.9 \\ 
  5 & 0.5 & 0.5 \\ 
  6 & 0.4 & 0.4 \\ 
  7 & 0.3 & 0.3 \\ 
  8 & 0.2 & 0.2 \\ 
  9 & 0.9 & 0.9 \\ 
  10 & 0.6 & 0.6 \\ 
  11 & 0.7 & 0.7 \\ 
  12 & 0.8 & 0.8 \\ 
  13 & 0.1 & 0.1 \\ 
   \hline
\end{tabular}
\label{tab:simdata}
\end{table}

\begin{table}[ht]
\centering
\caption{Specified parameter values for $\boldsymbol{\theta}$ in the non-binary Dean and Raftery simulated dataset for the discriminating variables.}
\begin{tabular}{rrrrr}
  \hline
Variable & Category & Class 1 & Class 2 & Class 3 \\ 
  \hline
1 & 1 & 0.1 & 0.3 & 0.6 \\ 
   & 2 & 0.1  & 0.5  & 0.2 \\ 
   & 3 & 0.8 & 0.2 & 0.2 \\ 
  2 & 1 & 0.5 & 0.1 & 0.7\\ 
   & 2 & 0.5 & 0.9 & 0.3\\ 
  3 & 1 & 0.2 & 0.7 & 0.2 \\ 
   & 2 & 0.2 & 0.1 & 0.6\\ 
   & 3 & 0.3 & 0.1 & 0.1\\ 
   & 4 & 0.3 & 0.1 & 0.1 \\ 
  4 & 1 & 0.1 & 0.6  & 0.4 \\ 
   & 2 & 0.5 & 0.1 & 0.4\\ 
   & 3 & 0.4 & 0.3 & 0.2\\ 
   \hline
\end{tabular}
\label{tab:simdata21}
\end{table}

\begin{table}[ht]
\centering
\caption{Specified parameter values for $\boldsymbol{\theta}$ in the non-binary Dean and Raftery simulated dataset for the non-discriminating variables.}
\begin{tabular}{rrrrr}
  \hline
Variable & Category & All classes \\ 
  \hline
 5  & 1 & 0.4 \\ 
    & 2 & 0.5 \\
   & 3  & 0.1 \\
6  & 1 & 0.2  \\
  & 2 & 0.4 \\
  & 3 & 0.1 \\
     & 4 & 0.3 \\
 7 & 1 & 0.2 \\
  & 2 & 0.3 \\
  & 3 & 0.3 \\
  & 4 & 0.1 \\
   & 5 & 0.1 \\
 8 & 1 & 0.2 \\
  & 2 & 0.8 \\
 9 & 1 & 0.7 \\
  & 2 & 0.1 \\
      & 3 & 0.2 \\
  10 & 1 & 0.1 \\
  & 2 & 0.2 \\
  & 3 & 0.1 \\
  & 4 & 0.6 \\
   \hline
\end{tabular}
\label{tab:simdata22}
\end{table}

% \begin{table}[ht]
% \centering
% \caption{Specified parameter values for $\boldsymbol{\theta}$ in the Dean and Raftery simulated dataset.}
% $\mathbb{P}(X_{nm} = 1 \mid Z_{ng} = 1)$
% \begin{tabular}{rrrrrrrrrrrrrr}
%   \hline
%  $m=$& 1 & 2 & 3 & 4 & 5 & 6 & 7 & 8 & 9 & 10 & 11 & 12 & 13 \\ 
%   \hline
% Group 1 & 0.6 & 0.8 & 0.7 & 0.6 & 0.5 & 0.4 & 0.3 & 0.2 & 0.9 & 0.6 & 0.7 & 0.8 & 0.1 \\ 
% Group 2 & 0.2 & 0.5 & 0.4 & 0.9 & 0.5 & 0.4 & 0.3 & 0.2 & 0.9 & 0.6 & 0.7 & 0.8 & 0.1 \\ 
%    \hline
% \end{tabular}
% \label{tab:simdata}
% \end{table}

\subsubsection*{Alzheimer dataset}
This dataset of patient symptoms was recorded in the Mercer Institute of St. James' Hospital in Dublin, Ireland \citep{Walsh04, Walsh06}. The data consist of a record of the presence or absence of $M=6$ symptoms displayed by $N=240$ patients diagnosed with early onset Alzheimer's disease, and are available in the {\bf R}\citep{R} package {\bf BayesLCA}\citep{BayesLCA}. Previous studies had difficulty in determining whether two or three groups are more suitable for the data, where fitting a three group model also created difficulties when performing inference \citep{Walsh06}.

\subsubsection*{Teaching styles dataset}
This dataset was collected in an attempt to ascertain the different types of teaching style being employed in schools in the United Kingdom in the mid 1970s, and was discussed at length by \citet{aitkin1981} after an initial analysis by \citet{bennet76}. The dataset records which of $M=39$ teaching methods are employed by the $N=467$ schools. Computational limitations at the time meant that only a two or three group clustering of the data were seriously considered, with dimension reduction in a previous study performed using principal component analysis. Further, use of information based criteria such as the BIC had not been developed at that time, so that no automatic comparison of the clusterings could be made, with a decision on the optimal clustering being based on careful consideration of the different properties of the clusterings.

\subsubsection*{Physiotherapy dataset}
We also apply our methods to a physiotherapy dataset based on a survey recently conducted by Keith Smart in St. Vincent's University Hospital, Dublin \citep{Smart2011}. The aim of this study was to identify which symptoms distinguish between three types of back pain which patients suffer from. The dataset consists of $N=425$ observations and $M=36$ variables. While the different types of back pain were considered reasonably distinct, different subgroups of pain sufferers within these pain classes are also possible, which motivates an examination of the data in an unsupervised setting.

\section{Bayesian latent class model and marginalization approach}\label{sec:collapse}
In this section, a variable inclusion indicator variable is introduced, and a fully Bayesian specification for LCA is provided. A collapsed sampling scheme for inference is then described.

\subsection{Clustering variable selection}

The marginalization approach addresses exploring uncertainty in variable selection for clustering the items as well as uncertainty in the number of classes, $G$. To this end, let $\boldsymbol{\nu}_{\mathrm{cl}}$ be a vector containing the indexes of the set of variables used for clustering the data and $\boldsymbol{\nu}_{\mathrm{n}}$ contain the remaining indexes of variables not used for clustering the data. To make this clearer, we formulate the data model by using two parts. For brevity, we use $\boldsymbol{\nu}$ to denote $(\boldsymbol{\nu}_{\mathrm{cl}},\boldsymbol{\nu}_{\mathrm{n}})$. The variables in $\boldsymbol{\nu}_{\mathrm{cl}}$ follow an LCA model with $G$ classes. This gives the likelihood
\begin{eqnarray*}
p_{\mathrm{cl}}(\bX,\bZ|\boldtheta,\boldnu,\boldtau,G) &=& \prod^N_{n=1} \prod_{g=1}^G \left \lbrace \tau_g \prod_{m \in \nu_{\mathrm{cl}}} \right.\\
&\times& \left .  \prod_{c=1}^{C_m} \theta_{gmc}^{\indicator(X_{nm} = c)} \right \rbrace^{Z_{ng}}.
\end{eqnarray*}
The second part of the model concerns $\boldsymbol{\nu}_{\mathrm{n}}$. The variables in this index vector can still be seen to follow an LCA model, independent of that above. However, this time, the LCA model has just one class. This is since these variables do not hold information on the clustering. This gives the likelihood
\[
p_{\mathrm{n}}(\bX|\boldrho,\boldnu) = \prod_{n=1}^N \prod_{m \in \nu_{\mathrm{n}}} \prod_{c=1}^{C_m} \rho_{mc}^{\indicator(X_{nm} = c)},
\]
where $\rho_{mc}$ is the probability of variable $m$ having category $c$.

% The variables which are not used for clustering should follow the same distribution for all items, so that the data likelihood for these variables is
%\[
%p_{\mathrm{n}}(\bX|\boldrho,\boldnu) = \prod_{n=1}^N \prod_{m \in \nu_{\mathrm{n}}} \prod_{c=1}^{C_m} \rho_{mc}^{\indicator(X_{nm} = c)},
%\]
%where $\rho_{mc}$ is the probability of variable $m$ having category $c$.
%Similarly, the completed data likelihood for the clustering variables is
%\begin{eqnarray*}
%p_{\mathrm{cl}}(\bX,\bZ|\boldtheta,\boldnu,\boldtau,G) &=& \prod^N_{n=1} \prod_{g=1}^G \left \lbrace \tau_g \prod_{m \in \nu_{\mathrm{cl}}} \right.\\
%&\times& \left .  \prod_{c=1}^{C_m} \theta_{gmc}^{\indicator(X_{nm} = c)} \right \rbrace^{Z_{ng}}.
%\end{eqnarray*}

\subsection{Prior assumptions and joint posteriors} \label{sec:priors}

Prior assumptions for all sets of multinomial probabilities as well as the component weights $\boldtau$ are independent Dirichlet distributions. For example, for the clustering variables $m\in \boldsymbol{\nu}_{\mathrm{cl}}$
\[
p(\boldtheta_{gm}|\beta) = \frac{\Gamma\left(C_m \beta\right)}{\Gamma\left(\beta\right)^{C_m}} \prod_{c=1}^{C_m} \theta_{gmc}^{\beta-1}.
\]
The hyperparameters $\beta$ are chosen to be 1 for all $g,m$ combinations. A Dirichlet prior is also assumed for those variables $m \in \nu_{\mathrm{n}}$, where the hyperparameters $\beta$ are again chosen to be 1 in all cases:
\[
p(\boldrho_m | \beta) = \frac{\Gamma\left(C_m \beta\right)}{\Gamma\left(\beta\right)^{C_m}} \prod_{c=1}^{C_m} \rho_{mc}^{\beta-1}.
\]
For the component weights the prior assumed is 
\[
p(\boldtau|\alpha, G) = \frac{\Gamma\left(G \alpha\right)}{\Gamma\left(\alpha\right)^G} \prod_{g=1}^G \tau_g^{\alpha-1}
\]
where $\alpha$ is taken to be 0.5, for all possible values of $g$; this is the marginal Jeffreys for the multinomial model, and is partly chosen to discourage overfitting \citep{Rousseau2011}.

It is assumed that the prior probability of any variable being a clustering variable is $\pi$, giving a joint prior for $\boldnu$ of 
\[
p(\boldnu|{\pi}) = \prod_{m\in \nu_{\mathrm{cl}}} \pi \prod_{m \in \nu_{\mathrm{n}}}(1-\pi).
\]
This prior is independent of the priors on $\boldtheta,\boldrho$ and $\boldsymbol{\tau}$.
~\citet{ley2009} give a detailed discussion of this prior on variable inclusion. They conclude that further assuming a $\mbox{Beta}(a_0,b_0)$ hyperprior on $\pi$ can be much more sensible than using a fixed value of $\pi$. We investigate the use of such a hyperprior in analyzing the Alzheimer data in Section~\ref{sec:alzheimer}. When a hyperprior is not used,  a value of $\pi=0.5$ is assumed for the applications in this paper.

The complete data likelihood for all variables is
\[
p_{\mathrm{full}}(\bX, \bZ |\boldtheta,\boldrho,\boldnu,\boldtau,G) = p_{\mathrm{cl}}(\bX, \bZ|\boldtheta,\boldnu,\boldtau,G) p_{\mathrm{n}}(\bX|\boldrho,\boldnu)
\]
and the joint posterior for the unknowns (except the number of classes) is
\begin{eqnarray*}
p(\bZ,\boldtheta,\boldrho,\boldnu,\boldtau|\bX,G, \alpha, \pi,\beta) &\propto& p_{\mathrm{full}}(\bX, \bZ | \boldtheta,\boldrho,\boldnu,\boldtau,G)\\  & \times & p(\boldtau|\alpha, G) p(\boldnu | \pi)\\
 & \times & \prod_{m \in \nu_{\mathrm{n}}} p(\boldrho_m | \beta)\\
 & \times & \prod_{g=1}^G \prod_{m \in \nu_{\mathrm{cl}}} p(\boldtheta_{gm} | \beta).
\end{eqnarray*}

Under the model being considered there is also uncertainty in the number of classes. This is accounted for by taking a prior on $G$, which we assume to be $\mathrm{Poisson}(1)$, truncated to $\{1,\dots,G_{\mathrm{max}}\}$, with $G_{\mathrm{max}}$ the maximum number of classes considered feasible. We choose this prior for $G$ following its justification by~\citet{Nobile2005} in the analysis of Bayesian finite mixtures. The full posterior is then
\begin{eqnarray*}
& & p(G,\bZ,\boldtheta,\boldrho,\boldnu,\boldtau|\bX, \alpha, \pi,\beta )\\
&  & \qquad  \propto  p(\bZ,\boldtheta,\boldrho,\boldnu,\boldtau|\bX,G,  \alpha, \pi,\beta) p(G).
\end{eqnarray*}
Note that this derivation of the posterior assumes that $\pi$ is a fixed value. If we include the hyperprior on $\pi$ discussed above, the right hand side of this proportionality relation has an extra multiplicative term $p(\pi) = \mbox{be}(\pi;a_0,b_0)$, where $\mbox{be}(\cdot;a_0,b_0)$ is the beta density with shape parameters $a_0$ and $b_0$.

\subsection{Marginalization approach}

The marginalization approach proceeds by observing that all of the multinomial probabilities, as well as the component weights can be marginalized analytically out of the model by using the normalizing constant of the Dirichlet distribution. This leaves a joint distribution for $G$, $\boldnu$ and $\bZ$:
\begin{eqnarray*}
 & ~ &p(G,\bZ,\boldnu|\bX, \alpha, \pi,\beta)\\
 &\propto& p(G) \int p(\bZ,\boldtheta,\boldrho,\boldnu,\boldtau|\bX,G, \alpha, \pi,\beta)\, d \boldtheta \, d \boldrho \, d \boldtau.
\end{eqnarray*}
Carrying out this marginalization gives
\begin{eqnarray}
& & p(G,\bZ,\boldnu|\bX, \alpha, \pi,\beta) \nonumber \\
& \propto & p(G)p(\boldnu|\pi)\frac{\Gamma\left(G \alpha\right)}{\Gamma\left(\alpha\right)^{G}}  \frac{\prod_{g=1}^G \Gamma\left(N_g + \alpha\right)}{\Gamma\left(N + G \alpha\right)} \nonumber \\
& \times &
\prod_{m \in \nu_{\mathrm{n}}}\frac{\Gamma\left(C_m \beta\right)}{\Gamma\left(\beta\right)^{C_m}} \frac{\prod_{c=1}^{C_m} \Gamma \left(N_{mc} + \beta\right)}{\Gamma \left(N + C_m \beta\right)} \nonumber \\
& \times & \prod_{g=1}^G \prod_{m \in \nu_{\mathrm{cl}}} \frac{\Gamma\left(C_m \beta\right)}{\Gamma\left(\beta\right)^{C_m}}\frac{\prod_{c=1}^{C_m}\Gamma\left(N_{gmc}+\beta\right)}{\Gamma\left(N_g + C_m \beta\right)} \label{eq:fpost}
\end{eqnarray}
where $N_g$ is the number of observations clustered to group $g$, $N_{mc}$ is the number of times variable $m$ takes category $c$, and $N_{gmc}$ is the number of items in group $g$ that have category $c$ for variable $m$.
Of note here is that this posterior makes sampling the number of groups $G$ and the clustering variables $\boldsymbol{\nu}_{\mathrm{cl}}$ possible using standard MCMC techniques as outlined in the following section. 

\subsection{Sampling algorithm} \label{sec:sampling algorithm}

The sampling algorithm comprises three main operations. The first samples the class membership of observations, the second samples the number of classes and the third step samples the variables useful for clustering.

\subsubsection{Class memberships}\label{sec:class:memberships}

Class memberships are sampled using a Gibbs sampling step which exploits the full conditional distribution of the class label for observation $n$, $n=1,\dots,N$. Given the current configuration of labels, groups and clustering variables, and supposing that the current $Z_{ng} = 1$, we can find the full conditional probability of item $n$ belonging to any class $h \ne g$, by taking it out of class $g$ and putting in this class in (\ref{eq:fpost}). The full conditional of remaining in class $g$ is just (\ref{eq:fpost}). The values for $h\ne g$ can be normalized by the case where the class remains the same, so that up to a normalizing constant, the full conditional distribution of $\bZ_n$ is found by: %through the relation

\begin{enumerate}[(a)]

\item evaluating Equation (\ref{eq:fpost}) $G$ times, at the current values of $\lbrace  G, \bZ \backslash  \bZ_n, \nu \rbrace$;
 \item setting $Z_{ng}$ to be proportional to its corresponding value, for $g  = 1\ldots, G$; 
\item normalizing the resulting values, so that \\$\sum^G_{g=1} Z_{ng} = 1.$

\end{enumerate}
%\begin{eqnarray*}
%& ~ & p(Z_{ng}=1|G,\boldnu, \bX, \alpha, \pi,\beta) \propto 1 \\
%& ~ & \qquad \qquad \mbox{and} \qquad \\
%&& p(Z_{nh}=1| G,\boldnu, \bX, \alpha, \pi,\beta) \propto  \\
%& & \\
%& ~&   \frac{ \Gamma\left(N_g-1+\alpha\right) \Gamma\left(N_h+1+\alpha\right) }{\Gamma\left(N_g+\alpha\right) \Gamma\left(N_h+\alpha\right)}\\
%& \times & \prod_{m \in \nu_{\mathrm{cl}}} \prod_{c=1}^{C_m}\frac{ \Gamma\left(N_{gmc} - \mathrm{I}(X_{nm} = c) + \beta \right) }{\Gamma\left(N_{gmc} + \beta \right)}
%\\ & & \qquad \qquad \qquad \times \frac{\Gamma(N_g + C_m \beta)}{\Gamma\left(N_g-1+C_m\beta\right) }
%\\ 
%&\times &  \prod_{m \in \nu_{\mathrm{cl}}}\prod_{c=1}^{C_m} \frac{\Gamma\left(N_{hmc} + \mathrm{I}(X_{nm} = c) + \beta\right)}{\Gamma\left(N_{hmc} + \beta\right)}\\
%& &\qquad \qquad \qquad \times \frac{\Gamma\left( N_h + C_m\beta\right)}{\Gamma\left(N_h + 1 + C_m \beta\right)},
%\end{eqnarray*}
%for $h\neq g.$

Memberships are updated for each observation at each iteration. If we let $C = \sum_{m=1}^M C_m$, we can see by inspection from this, that the computational effort required to sample the class memberships in one sweep of the algorithm is in the worst case $O(2 N M C G )$. There are various scenarios when this can become prohibitive for long runs of the sampler. For example, we may always expect $C$ to not have too large a magnitude, as possibly with $G$ in some cases. If we have reason to believe only a fraction of the $M$ possible variables are relevant for clustering, the main cost comes from the number of samples $N$. 

\subsubsection{Number of classes}

The number of classes may be sampled using the approach introduced by \citet{nobile07} for Gaussian mixtures. We outline this approach here. Given that there are currently $G$ components, two Metroplis-Hastings moves may be proposed: either the observations assigned to a component are randomly divided into two groups, so that a new component is ``ejected'' from an existing one, and the number of groups is increased to $G+1$, or the observations in two distinct groups are merged together, so that a component is ``absorbed'', and the number of groups decreases to $G-1$.                

A component is added or removed with probability $p_G = 0.5$ or $1-p_G  = 0.5$, except at the endpoints $G \in \lbrace1, G_{\max}\rbrace,$ where they are modified appropriately. A component $k$ is chosen at random to ``eject'' a new component from. To eject the new component, a draw $u\sim \mathrm{Beta}(a,a)$ is made, and each element of the ejecting component $k$ is assigned to the new component $G+1$ with probability $u$, otherwise it remains in $k$. The choice of shape parameter $a$ can have a strong effect on sampler performance, and is most effective when close to empty components are proposed often. This choice is determined by the size of the ejecting component, and a suitable value may be obtained by, for example, numerical programming. We refer to Appendix A3 in \citet{nobile07} for further details. 

If the proposed components' quantities are denoted by a $\tilde{\cdot}$, then the acceptance probability of an eject is $\min(1,A)$ where
\begin{eqnarray*}
 A &=& \frac{p(G+1,\tilde{\bZ},\boldnu|\bX, \alpha, \pi,\beta)}{p(G,\bZ,\boldnu|\bX, \alpha, \pi,\beta)} \times \frac{1-p_{G+1}}{p_G}\\
 &\times& \frac{\Gamma\left(a\right)^2}{\Gamma\left(2a\right)} \times \frac{\Gamma\left(2a + N_k\right)}{\Gamma(a+\tilde{N}_{k})
  \Gamma(a + \tilde{N}_{G+1})}.
\end{eqnarray*}

If the move is accepted, an additional label swap between the ejected component and another of the components selected at random is carried out. This is to improve the mixing properties of the sampler. For the reverse absorption move, one chooses two components $k$ and $k^{\prime}$ at random, and places all items from $k$ into $k^\prime$, computing the acceptance probability as $\min(1,A^{-1})$. If accepted, the number of components decreases from $G+1$ to $G$.

\subsubsection{Clustering variables}

To sample the clustering variables an index $j$ is chosen randomly from $\{1,\dots,M\}$. If $j \in \boldsymbol{\nu}_{\mathrm{n}}$ it is proposed to move it to $\boldsymbol{\nu}_{\mathrm{cl}}$. Alternatively, if $j \in \boldsymbol{\nu}_{\mathrm{cl}}$, it is proposed to move it to $\boldsymbol{\nu}_{\mathrm{n}}$.

If the chosen $j \in \boldsymbol{\nu}_{\mathrm{n}} $, the acceptance probability of the proposal to move it to $\boldsymbol{\nu}_{\mathrm{cl}}$ is $\min (1,R)$ with
\begin{eqnarray*}
R &=& \frac{p(G,\bZ,\tilde{\boldnu} | \bX, \alpha, \pi,\beta)}{p(G,\bZ,\boldnu | \bX, \alpha, \pi,\beta)} \\
&=&  \left ( \frac{\Gamma\left(C_j \beta\right)}{\Gamma\left(\beta\right)^{C_j}} \right )^{G-1} \prod^G_{g=1} \frac{\prod_{c=1}^{C_j}\Gamma(N_{gjc}+\beta)}{\Gamma(N_g + C_m \beta) } \\
&\times &\left ( { \frac{\prod_{c=1}^{C_j} \Gamma (N_{jc} + \beta)}{\Gamma (N + C_j \beta)} } \right )^{-1} \times \left(\frac{\pi}{1-\pi} \right).
\end{eqnarray*}
In this expression $\tilde{\boldnu}$ is the proposed new allocation of clustering and non-discriminating variables. A similar calculation can be carried out to compute the acceptance probability when the chosen $j \in \boldsymbol{\nu}_{\mathrm{cl}}$.  

\subsubsection{Sampling $\pi$ with a $\mathrm{Beta}(a_0,b_0)$ hyperprior} \label{sec:hprior_pi}

Taking a $\mathrm{Beta}(a_0,b_0)$ hyperprior on $\pi$ as in~\citet{ley2009}, it can be shown that the full conditional of $\pi$ given everything else is $\mathrm{Beta}(|\boldsymbol{\nu}_{\mathrm{cl}}|+a_0, |\boldsymbol{\nu}_{\mathrm{n}}| + b_0)$, where $|\be|$ means the number of elements in $\be$. This can be sampled in each sweep using a draw from the beta distribution.

\subsubsection{Sampling scheme summary}

Each sweep of the algorithm thus takes the following three steps:
\begin{enumerate}[(a)]
\item Update the class membership of observations using a Gibbs sampling step.
\item Propose to add a component with probability $p_G$, otherwise propose to absorb (remove) a component.
\item Choose one variable at random. If it is not included to cluster variables, propose to do so. If currently included as a clustering variable, propose to exclude it.
\end{enumerate}
The iterations used for posterior inference are taken after an initial burn-in phase. In addition to the moves above, if one has assumed a hyperprior on $\pi$, a ``(d)'' step involves sampling from the full conditional in Section~\ref{sec:hprior_pi}. In the applications described in Section~\ref{sec:apply}, diagnostic tools in the {\bf{R}} package {\bf{coda}}~\citep{coda} were applied to the log posterior to help determine whether a sufficient number of samples have been run for burn-in to occur, and whether thinning of the resulting samples is required. Visually, trace plots of the log posterior, number of groups and number of variables included can also be used to assess the effectiveness of the sampler. 

%It may sometimes be useful to sub-sample the resulting chains to reduce correlation between samples, however here we use the entire chain for estimation of posterior quantities.

\section{Post-hoc procedures for inference}\label{sec:post-hoc}

The MCMC output from running the algorithm described  in Section~\ref{sec:sampling algorithm} can be post-processed in order to perform inference on the clustering of observations, as well as item probability and weight  parameter estimation. The first step is to correct the output samples of class labels for label switching. How this is done is outlined in the following. Note that no label switching is required in order to evaluate the number of groups  which underly the data, or which variables are useful for clustering. After this, other post-hoc procedures for inference are discussed. 

\subsection{Label switching}

Label switching can occur in the algorithm of Section~\ref{sec:sampling algorithm}. The reason is that 
\[
p(G,\bZ,\boldnu|\bX, \alpha, \pi,\beta) = p(G,\bZ_{\cdot \delta},\boldnu|\bX, \alpha, \pi,\beta)
\]
where $\bZ_{\cdot\delta}$ denotes the indicator matrix obtained by applying any permutation $\delta$ of $1,\dots,G$ to the columns in $\bZ$. This invariance to permutations of the labels makes posterior inference of the clusterings fruitless unless some post-processing procedure is employed to try to ``undo'' the label switching first \citep{stephens2000,celeux2000,marin05}. Here, the procedure used is the same as that proposed by \citet{nobile07} and discussed in detail in \citet{wyse10}. We provide  a brief outline in the following. 

The method re-labels samples by minimising a cost function of the group membership vectors $\mathbf{Z}$. Let $\mathbf{Z}^{(T)}$ denote the value of the group membership indicator matrix $\mathbf{Z}$ stored at iteration $T$ during the sample run. Then a $G \times G$ cost matrix $\mathbf{C}$ can be created with entries
$$C_{gh} =  \sum^{T-1}_{t=1}\sum^N_{n=1} (1 - Z^{(t)}_{ng}){Z}^{(T)}_{nh}.$$
% $$\mathbf{C} =  \sum^{T-1}_{t=1}\left({\bZ^{(t)}\right )}^{\top}{\bZ}^{(T)}.$$
A cost function for $\mathbf{C}$ can then be constructed which is minimised by the permutation $\delta$ of $\mathbf{Z}^{(T)}_{n}$, $n=1, \ldots, N,$ which minimises the trace of $\mathbf{C}$. This is found using the square assignment algorithm of \citet{Carpaneto1980}.

\subsection{Post-hoc parameter estimation}
Although we integrate out the mixture weight and item probability parameters $\boldtau$ and $\boldtheta$ from the model, it is still possible to estimate $\emph{a posteriori}$ summaries of the parameters. Here we demonstrate how it is possible to obtain parameter expectation and variance estimates, conditional on a given number of groups, from a post-hoc calculation, by making use of the following formulae:
\begin{eqnarray*}
{\mathbb{E}}[A] &=& {\mathbb{E}}[{\mathbb{E}}[A | B] ] \\%\label{eq:Exp}\\
{\mathbb{V}}\mbox{ar}[A] &=& {\mathbb{E}}[{\mathbb{V}}\mbox{ar}[A | B] ]  + {\mathbb{V}}\mbox{ar}[{\mathbb{E}}[A| B] ],% \label{eq:Var}
\end{eqnarray*}
%We can think of the latter equation as the ``average of the variances plus the variance of the averages".
for any random variables $A$ and $B$. Clearly, in providing these posterior estimates, we condition on a given set of variables in the model for all iterations. These calculations can thus be performed using an auxiliary run of the sampler which keeps only the clustering variables in the model ($\mbox{i.e.}$ no variable search). 

Define the following summary statistics, for $t \in 1, \ldots, T$ iterations:
\begin{eqnarray*}
N^{(t)}_g &:=& \sum^N_{n=1} Z^{(t)}_{ng} \\
S^{(t)}_{gmc} &:= & \sum^N_{n=1} Z^{(t)}_{ng}\mathrm{I}(X_{nm} = c).
%S^{C(t)}_{gm} &:= & \sum^N_{n=1} Z^{(t)}_{ng}(1-X_{nm}).
\end{eqnarray*}
%Now, my reasoning is a little shaky, but hopefully the gist of what follows is right.
Then we can estimate the  expected values 
\begin{eqnarray*}
{\mathbb{E}}[\theta_{gmc} | \bX, \beta] & = & {\mathbb{E}}[{\mathbb{E}}[ \theta_{gmc} | \bX, \bZ, \beta] ]  \\
& \approx & \frac{1}{T} \sum^T_{t=1}{\mathbb{E}}[\theta_{gmc} | \bX, \bZ^{(t)}, \beta],
\end{eqnarray*}
%Now, $ p(\theta_{gm} | \bX, \bZ^{(t)}) \propto \prod^N_{n=1} \left( \theta_{gm}^{X_{nm}} (1 - \theta_{gm})^{(1 - X_{nm})}\right)^{Z^{(t)}_{ng}},$
%and since it's easy enough to recognise this as a beta distribution, it follows that $${\mathbb{E}}[\theta_{gm} | X, Z^{(t)}] = \frac{S^{(t)}_{gm}}{N^{(t)}_g}.$$
and since $ p(\boldtheta_{gm} | \bX, \bZ^{(t)}, \beta)$ follows a Dirichlet distribution, 
$${\mathbb{E}}[\theta_{gmc} | \bX, \beta]  \approx  \frac{1}{T} \sum^T_{t=1} \frac{S^{(t)}_{gmc} + \beta}{N^{(t)}_g + C_m\beta}.$$
It is easy to show based on a similar calculation that 
$${\mathbb{E}}[\tau_{g} | \alpha]  \approx  \frac{1}{T} \sum^T_{t=1} \frac{N^{(t)}_{g} + \alpha}{N + G\alpha}.$$
To calculate the variances:
\begin{eqnarray*}
&~& {\mathbb{V}}\mbox{ar}[\theta_{gmc} |\bX, \beta]\\ &=& {\mathbb{E}}[{\mathbb{V}}\mbox{ar}[\theta_{gmc} | \bX,\bZ, \beta] ]  + {\mathbb{V}}\mbox{ar}[{\mathbb{E}}[\theta_{gmc} | \bX,\bZ, \beta] ] \\
& \approx & \frac{1}{T} \sum^T_{t=1}\frac{(S^{(t)}_{gmc} + \beta)(N^{(t)}_g + (C_m-1)\beta - S^{(t)}_{gmc})}{(N_g^{(t)} +C_m\beta)^2(N_g^{(t)} +C_m\beta + 1)}\\
&+& \frac{1}{T} \sum^T_{t=1} \left( \frac{S^{(t)}_{gmc} + \beta}{N^{(t)}_g + C_m\beta} -  \frac{1}{T} \sum^T_{t=1} \frac{S^{(t)}_{gmc} + \beta}{N^{(t)}_g + C_m\beta}   \right)^2,
\end{eqnarray*}
where we have again made use of the fact that \\$p(\boldtheta_{gm} | \bX,\bZ^{(t)}, \beta)$ follows a Dirichlet distribution to calculate the variance ${\mathbb{V}}\mbox{ar}[\theta_{gmc} | \bX,\bZ^{(t)}, \beta]$.

%Using the same reasoning as before, $${\mathbb{V}}\mbox{ar}[\theta_{gmc} | \bX,\bZ^{(t)}]= \frac{S^{(t)}_{gmc}(N^{(t)}_g - S^{(t)}_{gmc})}{(N_g^{(t)})^2(N_g^{(t)} + 1)}.$$
%This gives $$ {\mathbb{V}}\mbox{ar}[\theta_{gm} |X] \approx \frac{1}{T} \sum^T_{t=1} \frac{S^{(t)}_{gm}S^{C(t)}_{gm}}{(N_g^{(t)})^2(N_g^{(t)} + 1)} + \frac{1}{T} \sum^T_{t=1} \left( \frac{S^{(t)}_{gm}}{N^{(t)}_g} -  \frac{1}{T} \sum^T_{t=1} \frac{S^{(t)}_{gm}}{N^{(t)}_g}    \right)^2.$$ 

Similarly, \begin{eqnarray*}
 {\mathbb{V}}\mbox{ar}[\tau_{g} | \alpha] &\approx& \frac{1}{T} \sum^T_{t=1} \frac{(N^{(t)}_{g} + \alpha)(N - N^{(t)}_{g} + (G-1)\alpha)}{(N+ G\alpha)^2(N + G\alpha + 1)} \\ 
 &+& \frac{1}{T} \sum^T_{t=1} \left( \frac{N^{(t)}_{g} + \alpha}{N + G\alpha} -  \frac{1}{T} \sum^T_{t=1} \frac{N^{(t)}_{g} + \alpha}{N + G\alpha}    \right)^2. \end{eqnarray*}

\subsection{Summaries of the sampler output}

The sampler deals with model selection on two levels within LCA. Selection of the number of groups, $G$, and selection of the clustering variables. We now suggest some useful summaries of the output which will aid in choosing an optimal model resulting from the search conducted through the algorithms proposed in Section~\ref{sec:sampling algorithm}. 

The first summary centres around examination of the approximate posterior for $G$ as represented by the sampler output. Let ${\boldsymbol{\cal G}}$ be a $G_{\max} \times T$ indicator matrix, where $T$ denotes the total number of iterations which the sampler runs for. We define an entry of ${\boldsymbol{\cal G}}$ to be: 
$$ {\cal G}_{kt} =  \left \lbrace  \begin{array}{ll} 1 &\mbox{if the chain has $k$ groups at iteration $t$;} \\  0 & \mbox{otherwise.}\end{array} \right .$$
Then, approximately, the posterior probability of $k$ groups is given by
$
p_k = \sum_{t=1}^T \mathcal{G}_{kt}/T.
$
These quantities can be examined to quantify the posterior support for $G$ groups.

The second summary aims to simultaneously summarize the joint uncertainty in the number of groups and variable inclusion. We construct an $G_{\max} \times M$ coincidence matrix ${\boldsymbol{\cal C}}$ where each entry indicates the amount of time which the sampler spent in a certain number of groups and including a certain variable. It is calculated as follows. Use ${\boldsymbol{\cal V}}$ to denote an $M \times T$ indicator matrix, where 
$$ {\cal V}_{mt} =  \left \lbrace  \begin{array}{ll} 1 &\mbox{if $m \in \boldsymbol{\nu}_{\mathrm{cl}}^{(t)}$;} \\  0 & \mbox{otherwise.}\end{array} \right .$$ 
Then each entry of ${\boldsymbol{\cal C}}$ is given by
$${\cal C}_{km} =  \frac{\sum^T_{t=1}{\cal V}_{mt}{\cal G}_{kt}}{\sum^T_{t=1}{\cal G}_{kt}}$$
where we have normalised the entry ${\cal C}_{km}$ so that it denotes the proportion, rather than the total amount of time the sampler spent in a particular model space. In words, the approximate probability of inclusion of variable $m$ as a clustering variable, conditional on $k$ groups, is $\mathcal{C}_{km}$. This coincidence matrix can be visualised as in the plots in Section~\ref{sec:apply} with each coloured rectangle giving a heat colour to represent $\mathcal{C}_{km}$. The closer the colour is to red (as opposed to blue), the closer $\mathcal{C}_{km}$ is to 1, that is, the more likely $m$ is {\it a posteriori} to be a good clustering variable for a $k$ group LCA model.

While in theory the matrix summarises the behaviour of the sampler for the entire model space, in practice some regions will not be visited by the sampler, with the corresponding entries being omitted in what follows.

\section{Data Applications}\label{sec:apply}

 \begin{figure*}[t]
\begin{center}
\centering
\subfigure{ \includegraphics[width=0.32\textwidth]{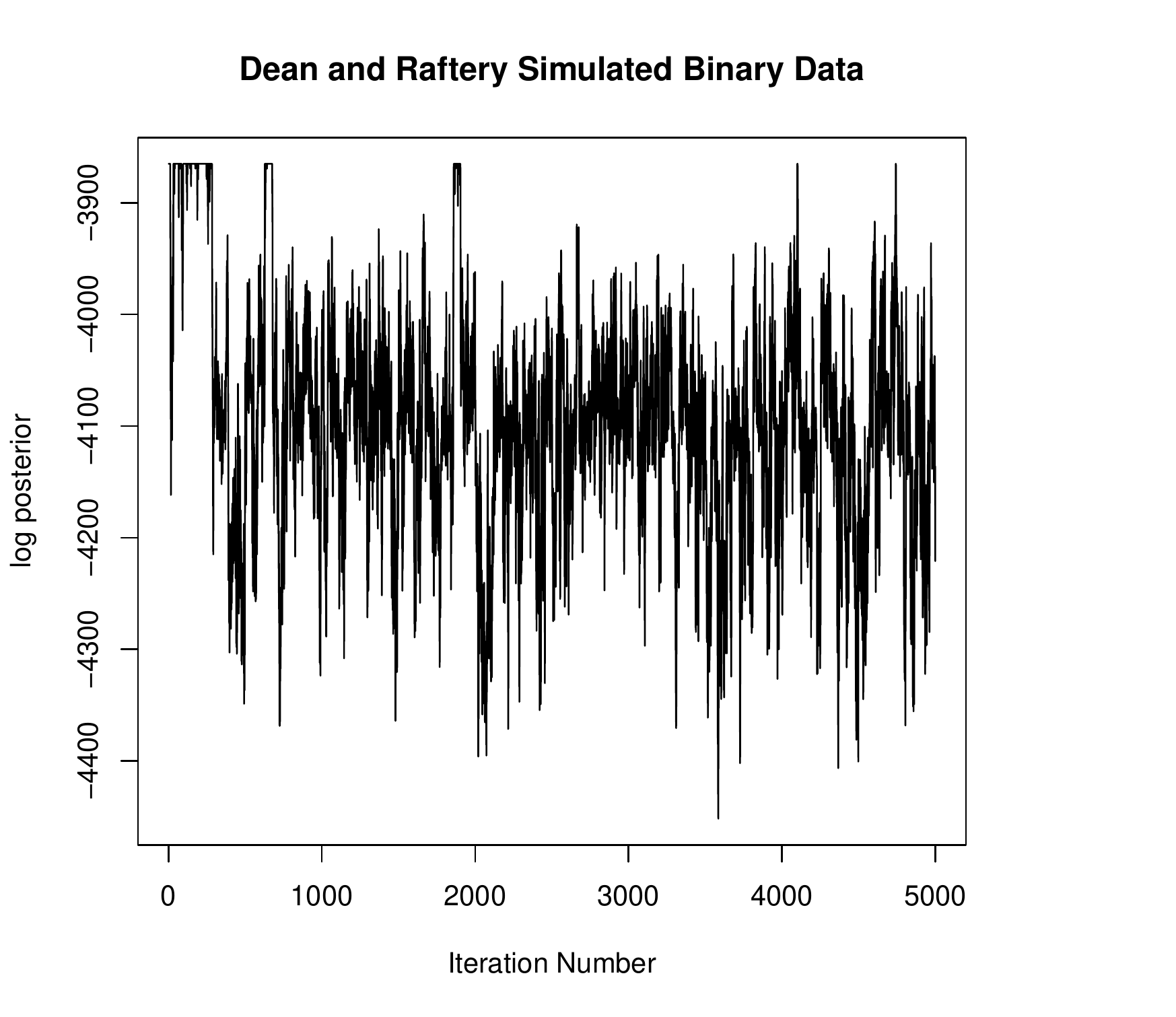}}
\subfigure{ \includegraphics[width=0.32\textwidth]{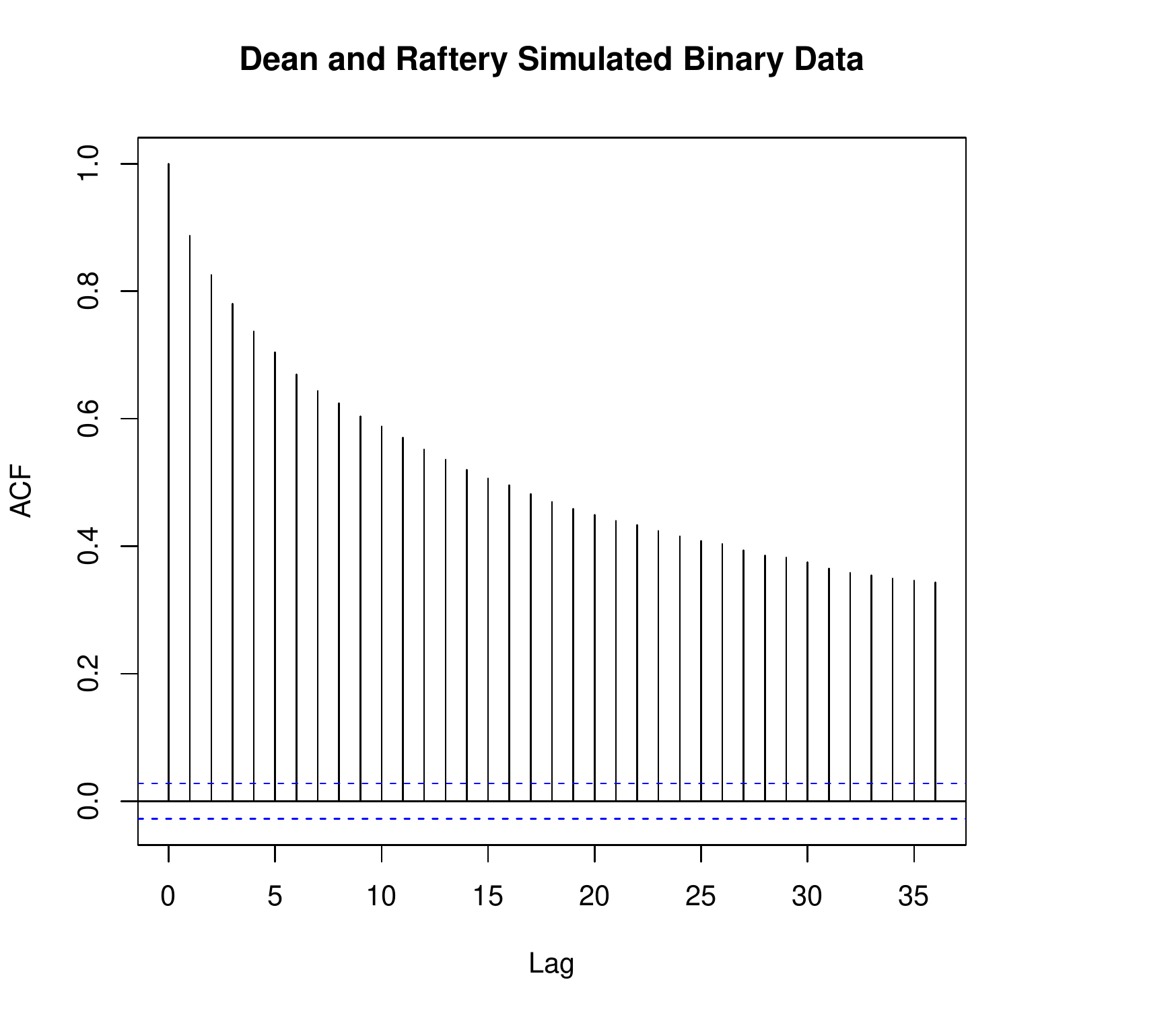}}
\subfigure{\includegraphics[width=0.32\textwidth]{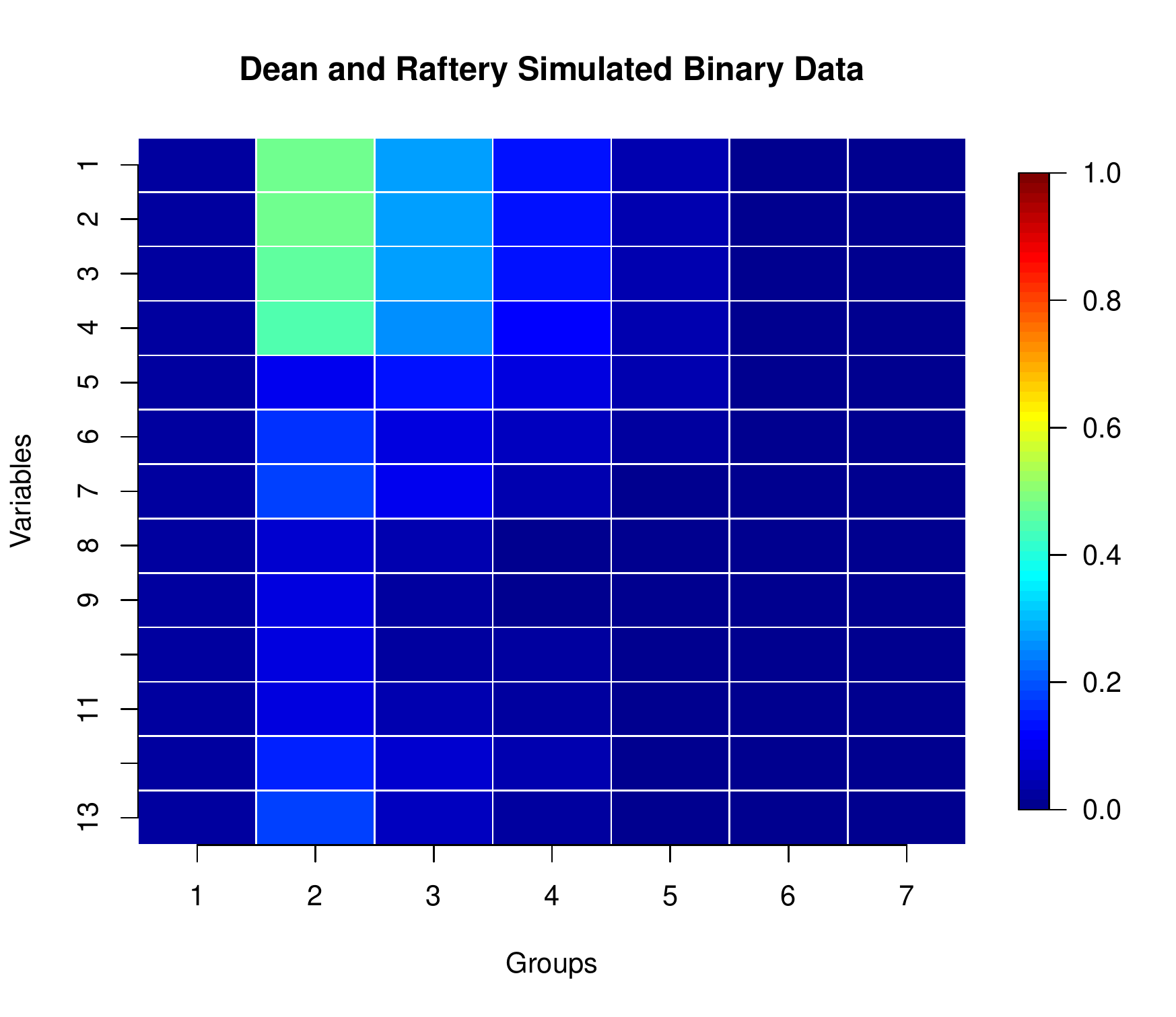}}
%\subfigure{\includegraphics[width=0.47\textwidth]{writeup_figs/Alzheimer_ngroups.pdf}}
%\subfigure{\includegraphics[width=0.47\textwidth]{writeup_figs/Alzheimer_nvariables.pdf}}
\end{center}
\caption{Diagnostic plots of the collapsed sampler applied to the simulated binary Dean and Raftery dataset, and on the right, a coincidence matrix for the sampler.  The model correctly identifies 2 groups as optimal, and frequently excludes the noise variables 5 -- 13.}
\label{fig:Sim}
\end{figure*}

%    \begin{figure*}[t]
% \subfigure{ 
% \includegraphics[width=0.48\textwidth]{Dean_Raftery_Coinc.pdf}
% \label{fig:Sima}
% }
% \subfigure{ 
% \includegraphics[width=0.48\textwidth]{DR_combined.pdf}
% \label{fig:Simb}
% } 
%  \caption{Summary of the spaces most commonly visited by the sampler over a run on simulated data. Right: a trace plot of the log-posterior and the number of groups from a run of the sampler on simulated data.}
%    \label{fig:Sim}
%    \end{figure*}
	
In this section the sampler is applied to the datasets described in Section~\ref{sec:data_description}. In all cases, a pilot run was first performed, before a better individually tuned sampler was then run on each dataset, based on the sampler's initial performance. Multiple runs of the sampler were also performed, to ensure consistency of results.

\subsection{Dean and Raftery data}

The algorithms described in Sections~\ref{sec:collapse} and~\ref{sec:post-hoc} were implemented in C and applied to the datasets in the R environment \citep{R}. Total run time varied depending on the size and particularly the number of variables of the dataset in question. For example, the total time taken to fit a model to the Alzheimer dataset, including post hoc parameter estimation, was about three and a half minutes. Fitting a finely tuned collapsed sampler to the larger Teaching dataset, without performing post hoc parameter estimation, took about 25 minutes, or about seven and a half times as long. Nevertheless, this should still be viewed as being reasonably efficient, as model and variable selection, as well as clustering, is simultaneously taking place.

%In order to visually summarise the sampler's behaviour and simultaneously identify which variables and number of groups are of most interest we construct an $M \times  G_{\max}$ coincidence matrix ${\boldsymbol{\cal C}}$. Each entry indicates the amount of time which the sampler spent for a certain group number and including a certain variable, and is calculated as follows.
%
%Let ${\boldsymbol{\cal G}}$ be a $G_{\max} \times T$ indicator matrix, where $T$ denotes the total number of iterations which the sampler runs for. We define an entry of ${\boldsymbol{\cal G}}$ to be: 
%$$ {\cal G}_{kt} =  \left \lbrace  \begin{array}{ll} 1 &\mbox{if the chain has $k$ groups at iteration $t$;} \\  0 & \mbox{otherwise.}\end{array} \right .$$
%Similarly, we use ${\boldsymbol{\cal V}}$ to denote an $M \times T$ indicator matrix, where 
%$$ {\cal V}_{mt} =  \left \lbrace  \begin{array}{ll} 1 &\mbox{if $\nu^{(t)}_m = 1$;} \\  0 & \mbox{otherwise.}\end{array} \right .$$ 
%
%Then each entry of ${\boldsymbol{\cal C}}$ is given by
%$${\cal C}_{km} =  \frac{\sum^T_{t=1}{\cal V}_{mt}{\cal G}_{kt}}{\sum^T_{t=1}{\cal G}_{kt}}$$
%where we have normalised the entry ${\cal C}_{km}$ so that it denotes the proportion, rather than the total amount of time the sampler spent in a particular model space. While in theory the matrix summarises the behaviour of the sampler for the entire model space, in practice some regions will not be visited by the sampler, with the corresponding entries being omitted in what follows. 

A sampler was run on the binary simulated data for 50,000 iterations after 1,000 iterations burn-in, with 1 in 10 samples retained by subsampling. The coincidence matrix is shown on the left hand side of Figure~\ref{fig:Sim}. This clearly identifies a two group model with variables 1-4 as being optimal for clustering.  In particular, variables 5-13 are included less than half the time, whereas variables 1-4 are included with high probability. The choice of two groups is decisively confirmed in Table~\ref{tab:DRgroup}, with a posterior probability of $0.4814$. 
%a Bayes factor of 18.9 in favour of two groups compared to one, and small Bayes factors for all other models. 
The sampler correctly classifies 381 of the 500 observations, only seven less than would be found using the true parameter values. These results are comparable to those found by \citet{dean10}.

\begin{figure}[t]
 \includegraphics[width=85mm]{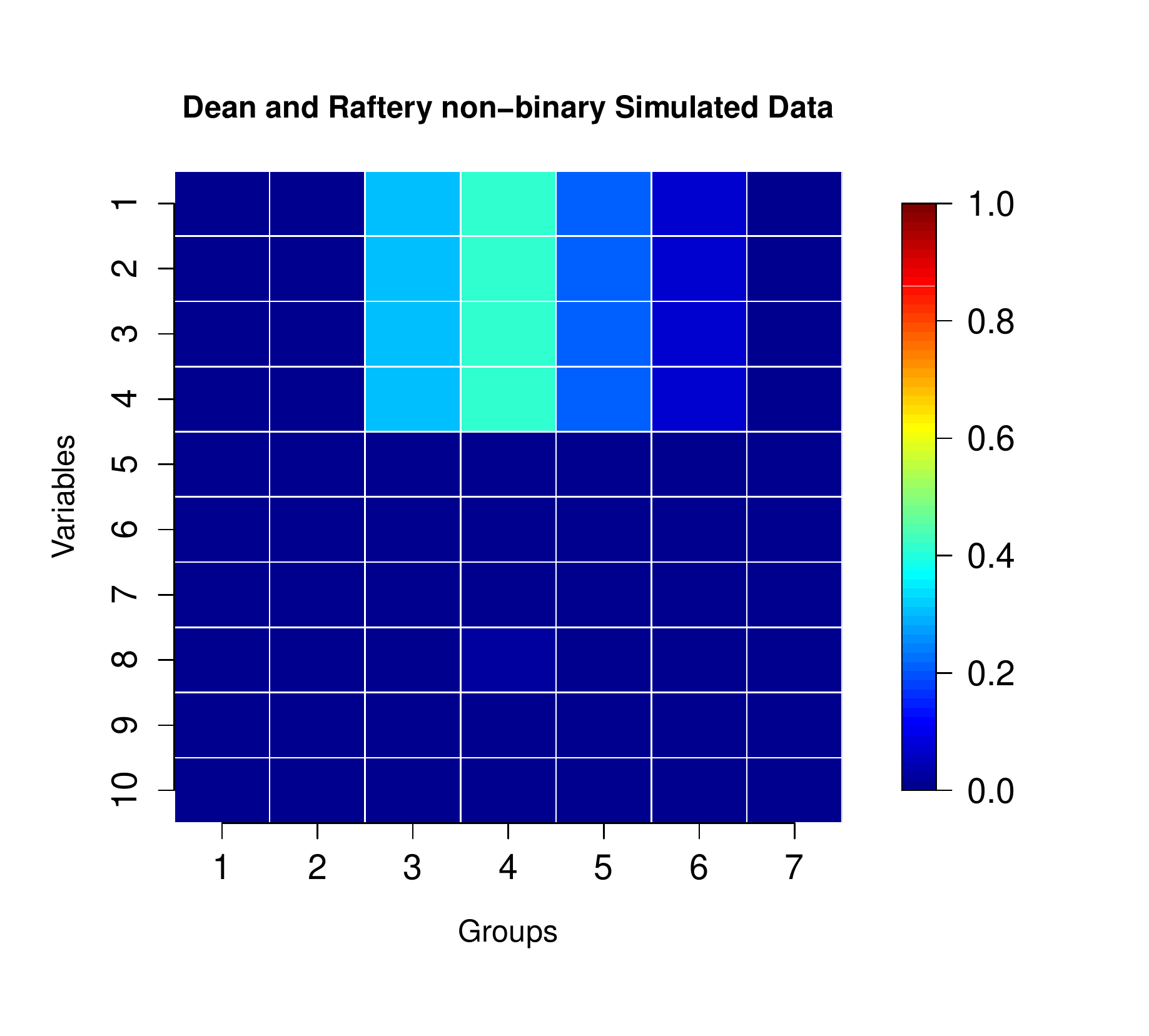}
  \caption{Summary of the spaces most commonly visited by the sampler over a run on the non-binary simulated data.}
    \label{fig:Sim2}
    \end{figure}

    \begin{table}[ht]

\caption{Posterior probability for number of groups in the binary Dean and Raftery data. A two group model is correctly identified as being optimal.}
\begin{tabular}{rrrrr}
  \hline
 $p_1$ &$p_2$ & $p_3$ & $p_4$ & $p_5$ \\ 
  \hline
  0.0510 & 0.4814 & 0.2840 & 0.1358 & 0.0396 \\ 
   \hline
\end{tabular}
\flushleft
\begin{tabular}{rr}
  \hline
 $p_6$& $p_7$ \\ 
  \hline
  0.0062 & 0.0020\\ 
   \hline
\end{tabular}
\label{tab:DRgroup}
\end{table} 
The sampler was run for 100,000 iterations after 10,000 burn-in for the non-binary data. The output was thinned and every 1 in 10 samples retained to construct the coincidence matrix shown in Figure~\ref{fig:Sim2}. Four classes is most likely {\it a posteriori} with $p_3 = 0.2986$, $p_4 = 0.4113$ and $p_5 = 0.2146$. The four informative variables (1-4) are the only ones which the sampler indicates are worth retaining.

\begin{table}[ht]

\caption{Posterior probability for number of groups in the non-binary Dean and Raftery data. }
\begin{tabular}{rrrrr}
  \hline
  $p_2$ & $p_3$ & $p_4$ & $p_5$ & $p_6$\\
  \hline
   0.0243  &  0.2986 & 0.4113 & 0.2146 & 0.0512  \\
   \hline
\end{tabular}
\label{tab:DRgroup2}
\end{table}

%    \begin{table}[ht]
%
%\caption{Posterior probability for number of groups in the non-binary Dean and Raftery data. }
%\begin{tabular}{rrrrrr}
%  \hline
%Method & $p_1$ & $p_2$ & $p_3$ & $p_4$ \\
%  \hline
%Collapsed Sampler &  & -  &  0.2986 & 0.4113 % \\  , .2986, .4113, .2146, 0.0512
%
%   Reversible Jump & 0.0924 & 0.2294 & 0.1414 & 0.1476 \\
%   \hline
%\end{tabular}
%\flushleft
%\begin{tabular}{rr}
%  \hline
%  $p_5$ & $p_{\geq 6}$\\ 
%  \hline
%   0.2146 & -  \\
%
%   0.1314 & 0.2578 \\ 
%   \hline
%\end{tabular}
%\label{tab:DRgroup2}
%\end{table}     

To investigate the effect of sample size on the sampler's performance we further simulated ten datasets each with $N = 1000, 2500, 5000, 10000$ from the non-binary model. Table~\ref{tab:N:insvest} shows the relative runtime (relative to $N=1000$), on a machine with 4 Gb RAM and 2.4GHz Intel i5 processor. The Rand index, based on how many would have been correctly predicted using the true parameter values, is also shown. It can be seen that the runtime increases roughly linearly with $N$ following the rough analysis in Section~\ref{sec:class:memberships}.

%    \begin{figure*}
% % Use the relevant command to insert your figure file.
% % For example, with the graphicx package use
%   \includegraphics[width=0.75\textwidth]{example.eps}
% % figure caption is below the figure
% \caption{Please write your figure caption here}
% \label{fig:2}       % Give a unique label
% \end{figure*}

%    \begin{table}[ht]
%\centering
%\caption{Bayes factors for group choice for the Dean and Raftery data. A two group model is correctly identified as being optimal.}
%\begin{tabular}{rrrrrrr}
%  \hline
% &$\mathcal{BF}_{1,2}$ &$\mathcal{BF}_{2,3}$ & $\mathcal{BF}_{3,4}$ & $\mathcal{BF}_{4,5}$ & $\mathcal{BF}_{5,6}$ &$\mathcal{BF}_{6,7}$\\ 
%  \hline
% & 18.88 & 1.77 & 1.91 & 1.46 & 0.94 & 2.26 \\ 
%   \hline
%\end{tabular}
%\label{tab:DRgroup}
%\end{table}
% 

\begin{table}
\caption{Runtimes and Rand indices for different simulated binary Dean and Raftery datasets of size $N$.} 
\begin{tabular}{rrr}
  \hline
 $N$ & Relative runtime & Rand index \\ 
  \hline
  1000 & 1  & 0.898 (0.045)\\
  2500 & 2.552  & 0.928 (0.023)\\
  5000 & 5.043  & 0.947 (0.032)\\
  10000 & 9.767  & 0.960 (0.017)\\ 
   \hline
\end{tabular}
\label{tab:N:insvest}
\end{table}

\subsection{Alzheimer Data} \label{sec:alzheimer}
\begin{figure*}[t]
\begin{center}
\centering
\subfigure{ \includegraphics[width=0.32\textwidth]{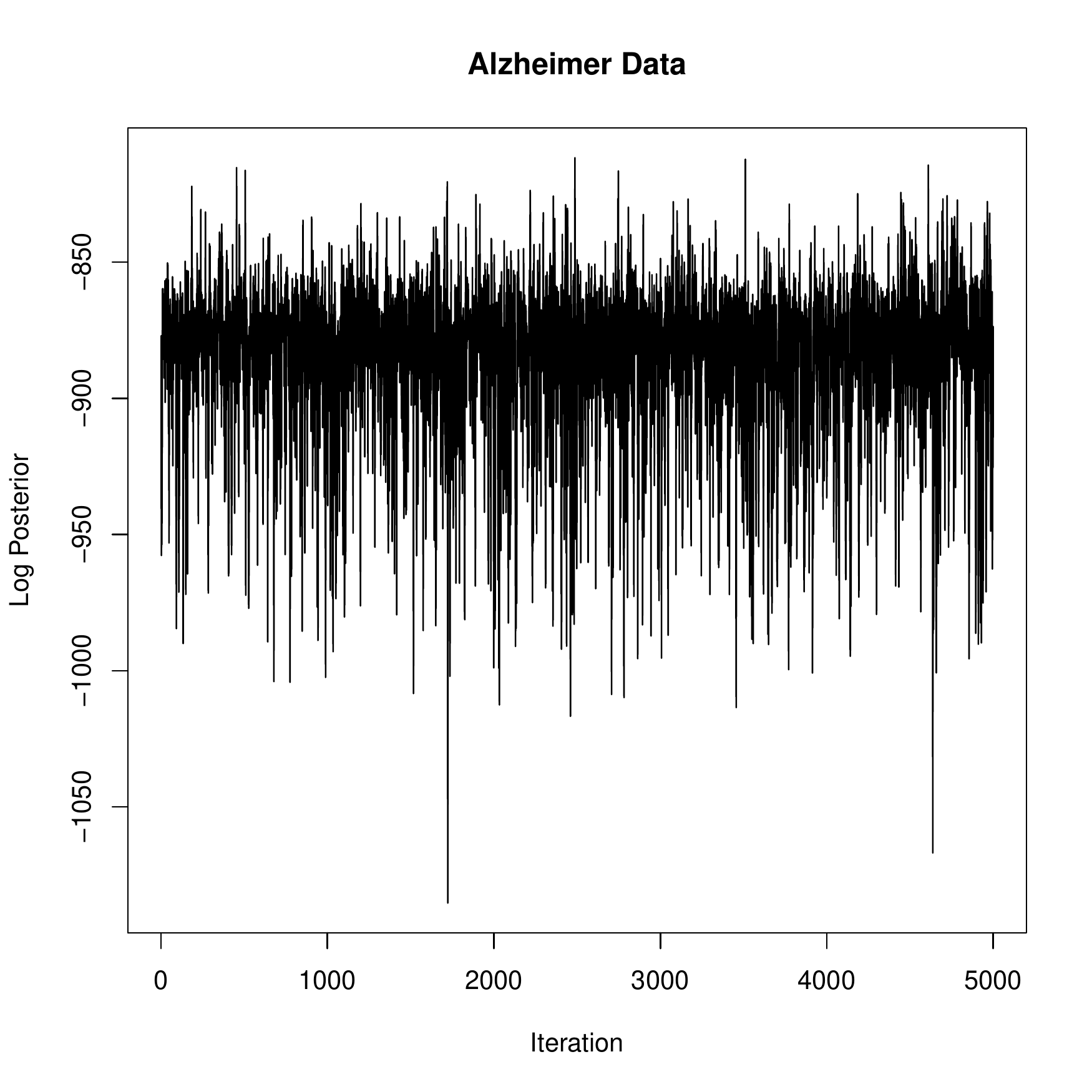}}
\subfigure{ \includegraphics[width=0.32\textwidth]{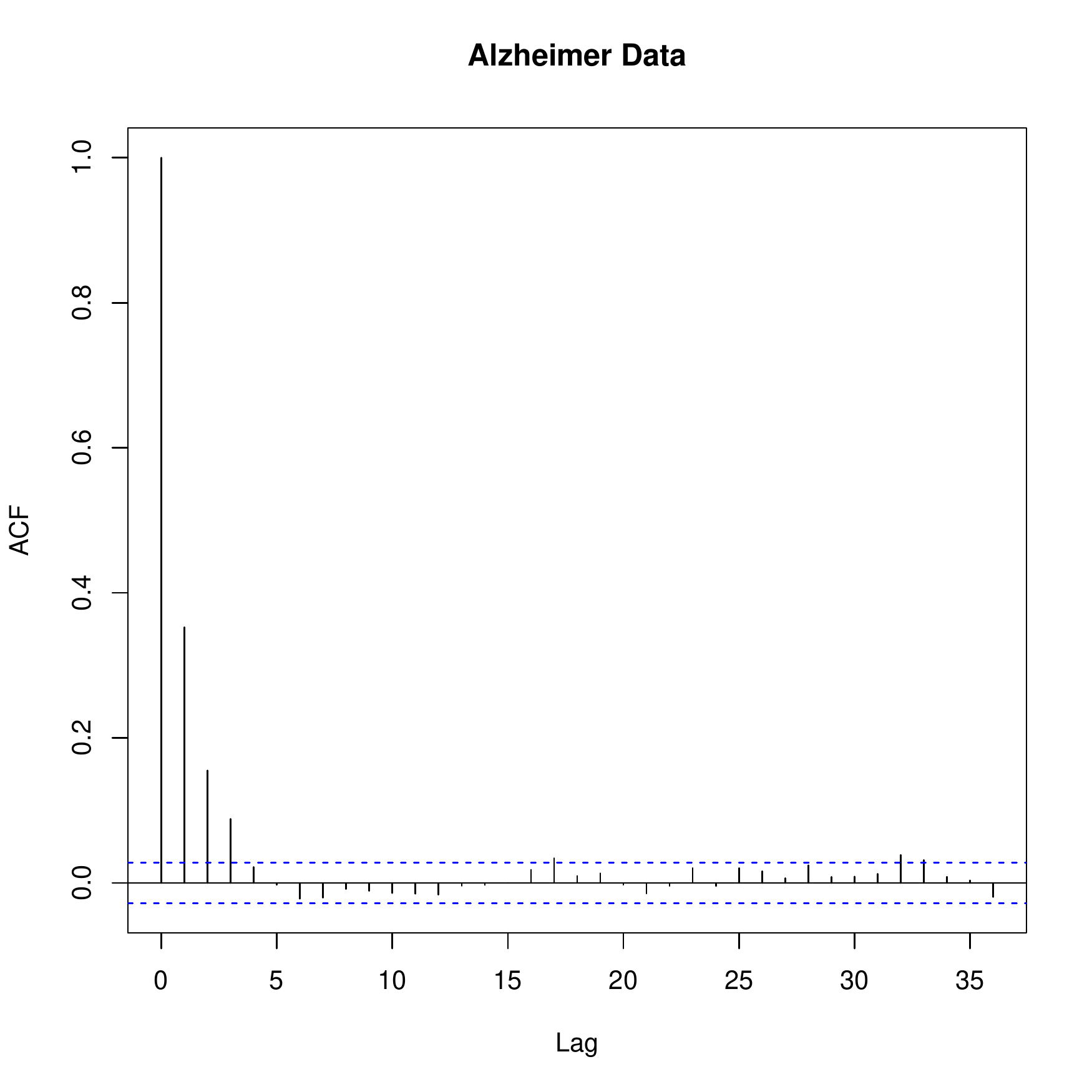}}
\subfigure{\includegraphics[width=0.32\textwidth]{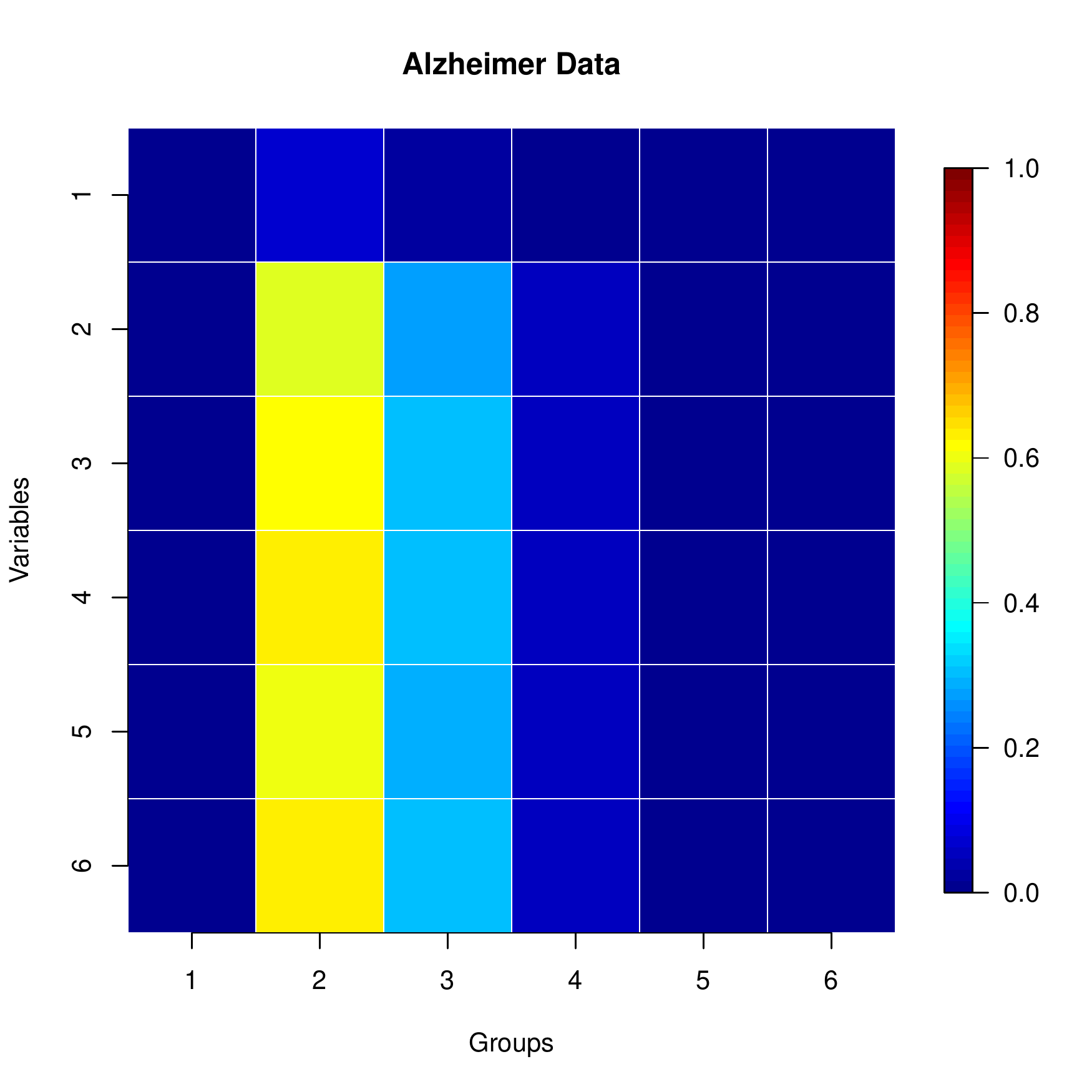}}
%\subfigure{\includegraphics[width=0.47\textwidth]{writeup_figs/Alzheimer_ngroups.pdf}}
%\subfigure{\includegraphics[width=0.47\textwidth]{writeup_figs/Alzheimer_nvariables.pdf}}
\end{center}
\caption{Diagnostic plots of the collapsed sampler applied to the Alzheimer dataset, and on the right, a coincidence matrix for the sampler.  The model appears to identify 2 groups as optimal, and frequently excluded variable 1 (Hallucination) during the run.}
\label{fig:Alzheimer_Eval}
\end{figure*}

Initially, the sampler was run on the Alzheimer data for 20,000 iterations after 1,000 iterations burn-in. While a visual inspection of the log posterior suggested that good mixing was occurring, the log posterior samples were found to have high autocorrelation, and diagnostic tools suggested that a longer sampling run was required. The sampler was then run for 100,000 iterations, and thinned by subsampling every twentieth iterate. While this substantially reduced the amount of autocorrelation between samples, the results of the clustering remained relatively unaffected. Diagnostic plots of the sampling run are shown in Figure~\ref{fig:Alzheimer_Eval}.

The coincidence matrix of the sampling run is shown in the rightmost plot of Figure~\ref{fig:Alzheimer_Eval}. The sampler excluded variable 1, Hallucination, a majority of the time, and identified 2 or 3 groups as optimal for the data, spending a majority of time in the 2 group space. The approximate posterior probability of a 2 group model is 0.6284, while that for a 3 group model is 0.2996 suggesting little evidence for a model with more than 2 groups. The approximate posterior of $G$ is given in Table~\ref{tab:alz:bf}.
%The Bayes factor for the evidence in favour of a 3 instead of group model 2 is $\mathcal{BF}_{2,3} = 1.43$, suggesting that there is little evidence in favour of the larger group choice. The Bayes factors for other group choices are smaller again. The Bayes factors for group choice are given in Table~\ref{tab:alz:bf}. The Bayes factors for variable selection are given in the supplementary material.

\begin{table}[thb]
\centering
\caption{Posterior probability for number of groups in Alzheimer data, firstly setting $\pi = 0.5$, and secondly using a $\mbox{Beta(1,1.5)}$ hyperprior on $\pi$.}
\begin{tabular}{rrrrrr}
  \hline
 Setting for $\pi$  & $p_2$ & $p_3$ & $p_4$ & $p_5$ &$p_6$ \\ 
  \hline
 $\pi = 0.5$ & 0.6284 & 0.2996 & 0.0622 & 0.0096 & 0.0002  \\ 
 $ \pi {\sim} \mbox{Beta(1,1.5)}$ &  0.6600 & 0.2724 & 0.0584 & 0.0092  & 0 \\ 
   \hline
\end{tabular}
\label{tab:alz:bf}
\end{table}

Additionally, we ran the sampler using a hyperprior on $\pi$ as outlined in Section~\ref{sec:priors} choosing hyperparameters $a_0 = 1$ and $b_0 = M/4 = 1.5$ as outlined in~\citet{ley2009}. The posterior for the number of classes is shown in the second row of Table~\ref{tab:alz:bf}. Hallucination (variable 1) was again excluded most of the time with a 2 class model being most preferred.  

%\begin{table}[htb]
%\centering
%\caption{Approximate posterior of $G$ for the Alzheimer data using a $\mbox{Beta(1,1.5)}$ hyperprior on $\pi$.}
%\begin{tabular}{rrrr}
%  \hline
%  $p_2$ & $p_3$ & $p_4$ & $p_5$  \\ 
%  \hline
%0.6600 & 0.2724 & 0.0584 & 0.0092  \\ 
%   \hline
%\end{tabular}
%\label{tab:alz:bf2}
%\end{table}

%\begin{table}[htb]
%\centering
%\caption{Bayes factors for group choice for the Alzheimer data.}
%\begin{tabular}{rrrrr}
%  \hline
% & $\mathcal{BF}_{2,3}$ & $\mathcal{BF}_{3,4}$ & $\mathcal{BF}_{4,5}$ & $\mathcal{BF}_{5,6}$ \\ 
%  \hline
% & 1.43 & 0.83 & 0.77 & 0.12 \\ 
%   \hline
%\end{tabular}
%\label{tab:alz:bf}
%\end{table}

We also compare parameter maximum \emph{a posteriori} and posterior standard deviation estimates for the item probability parameters of a 2 group model fitted to this dataset using the collapsed sampler to those obtained using a full Gibbs sampler in the {\bf BayesLCA} package. See \citet{garrett2000} and \citet{Walsh06} for descriptions of a full model Gibbs sampler method for LCA. %In the case where the included variables and number of groups are viewed as fixed, an appropriately tuned sampler can be used to estimate the posterior distribution of model parameters. 

Estimates from the full Gibbs sampler were obtained from 50,000 iterations after 1,000 iterations burn-in and, with every tenth sample retained. Estimates from the two methods are nearly identical. These are given in Table~\ref{tab:alz:par}. This suggests that while the most obvious uses of the collapsed sampler are to perform model and variable selection, as well as cluster the data, it can also be used as an effective tool for parameter estimation.  

\begin{table*}[t]
\centering
\caption{MAP item probability estimates (posterior standard deviations in parentheses) for the Alzheimer data obtained from post-hoc estimates from the collapsed Gibbs sampler and from a full model Gibbs sampler run. }
 \begin{tabular}{c}
Collapsed Gibbs sampler post-hoc estimates\\
\begin{tabular}{rrrrrrr}
  \hline
 & Hallucination & Activity & Aggression & Agitation & Diurnal & Affective \\ 
  \hline
Group 1 & 0.08 (0.03) & 0.54 (0.06) & 0.10 (0.04) & 0.14 (0.06) & 0.13 (0.05) & 0.59 (0.08)\\ 
  Group 2 & 0.10 (0.04) & 0.80 (0.06) & 0.40 (0.08) & 0.64 (0.12) & 0.39 (0.07) & 0.94 (0.04)\\ 
   \hline
\end{tabular} \\
~\\
Full model Gibbs sampler estimates\\
\begin{tabular}{rrrrrrr}
  \hline
 & Hallucination & Activity & Aggression & Agitation & Diurnal & Affective \\ 
  \hline
Group 1 & 0.08 (0.03) & 0.54 (0.06) & 0.11 (0.05) & 0.14 (0.06) & 0.14 (0.05) & 0.59 (0.08)\\ 
Group 2 & 0.10 (0.04) & 0.79 (0.07) & 0.39 (0.08) & 0.64 (0.12) & 0.38 (0.07) & 0.93 (0.07)\\ 
   \hline
\end{tabular}
\end{tabular}
\label{tab:alz:par}
\end{table*}

\subsection{Teaching Styles Data}
\begin{figure*}[t]
\begin{center}
\centering
\subfigure{ \includegraphics[width=0.32\textwidth]{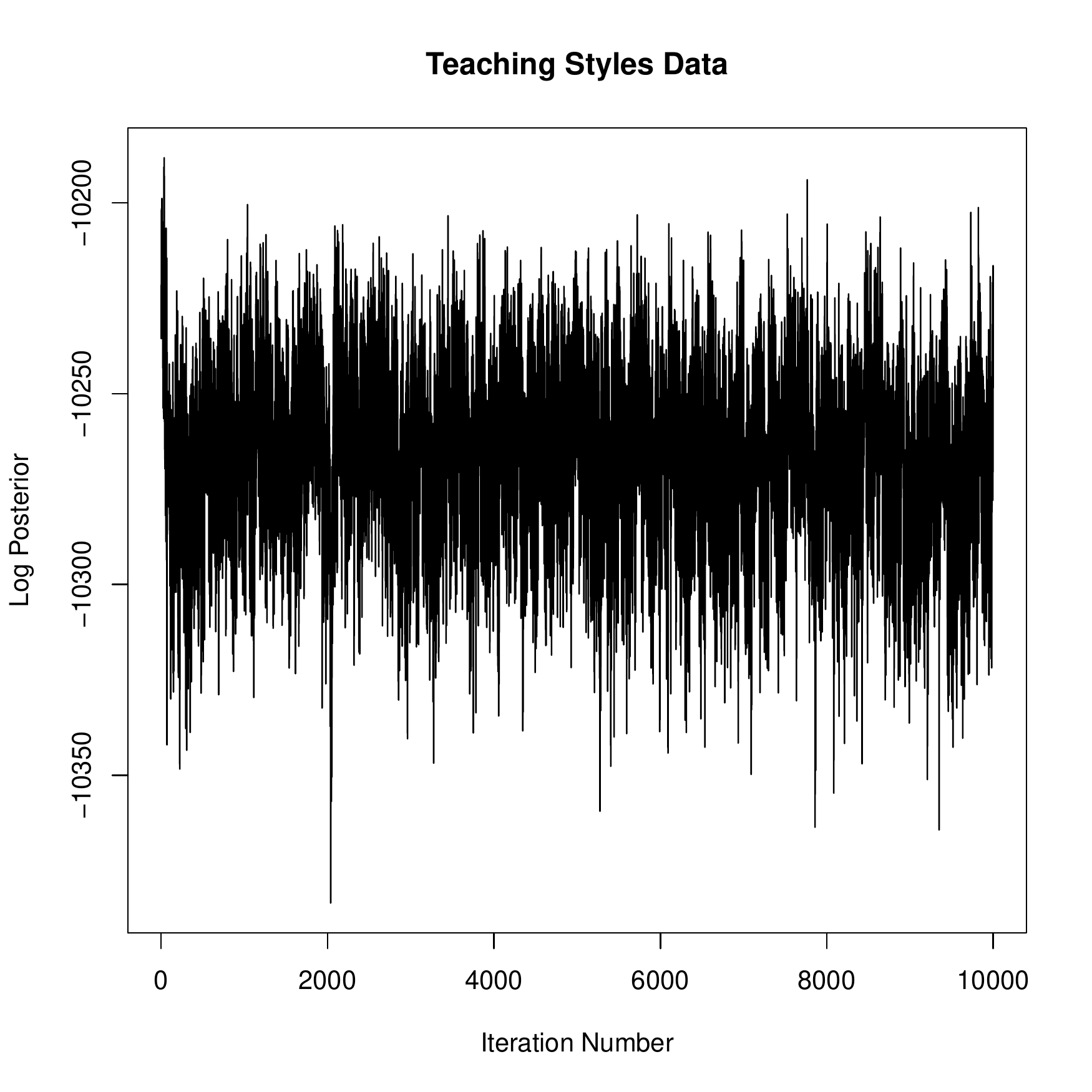}}
\subfigure{ \includegraphics[width=0.32\textwidth]{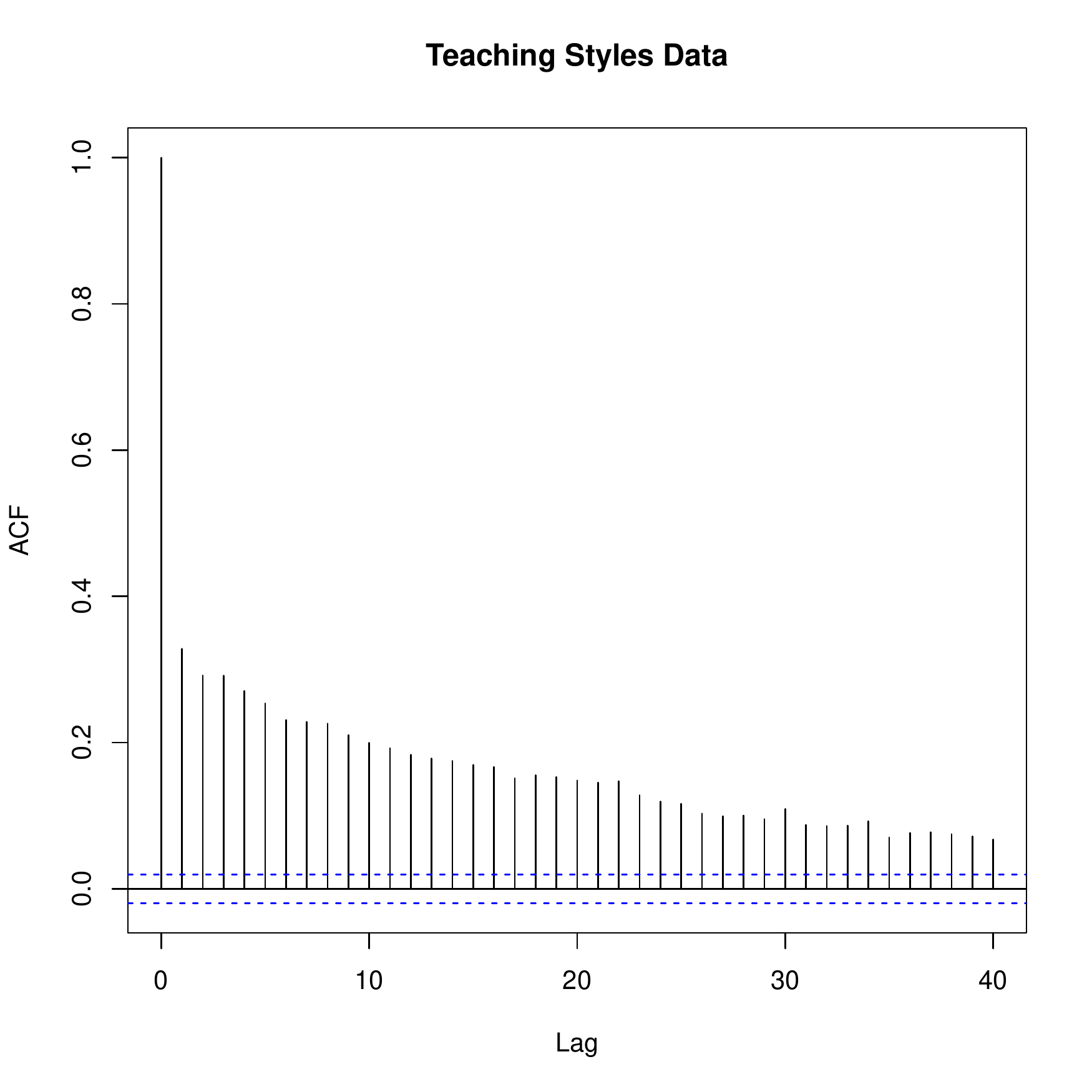}}
% \subfigure{\includegraphics[width=0.32\textwidth]{writeup_figs/Teach_LPngroups.pdf}}
% \subfigure{\includegraphics[width=0.32\textwidth]{writeup_figs/Teach_nvar.pdf}}
\subfigure{  \includegraphics[width=0.32\textwidth]{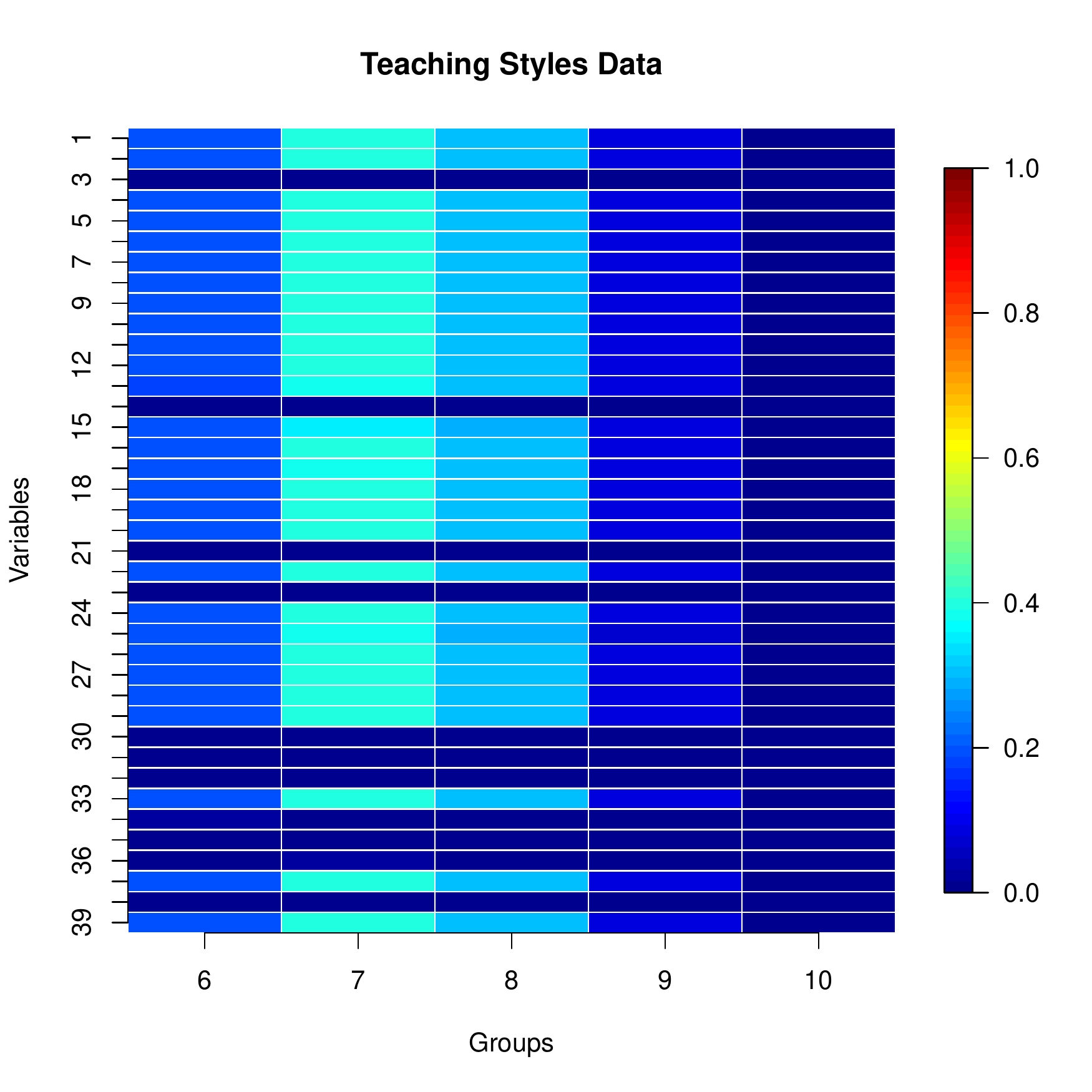}}
\end{center}
\caption{Diagnostic plots of the collapsed sampler applied to the Teaching Styles dataset, and on the right, a coincidence matrix for the sampler. }
% While the sampler doesn't appear to mix as well as in in the previous examples, it still explores the parameter space.}
\label{fig:Teach_Eval}
\end{figure*}

A collapsed sampler was run on the Teaching Styles data for 200,000 iterations after 5,000 iterations burn-in, with 1 in 20 samples retained by subsampling. Diagnostic tools suggest that the sampler behaves satisfactorily on the data. Diagnostic plots of the sampler are shown in Figure~\ref{fig:Teach_Eval}.

The coincidence matrix for the sampler is also shown as the rightmost plot in Figure~\ref{fig:Teach_Eval}. While several variables are decisively dropped by the sampler, a certain amount of uncertainty exists when it comes to selecting an optimal number of groups. Posterior probabilities for the number of groups are given in Table~\ref{tab:teach:group}. These suggest most evidence for a 7 group model, with lesser (but fair) support for 6, 8 and even 9 groups. Inspection of the clusterings shows that both the 7 and 8 group models cluster some observations into quite small groups; one group in the 7 group model has only 10 observations, while the two smallest groups in the 8 group model contain only 2 and 15 observations respectively. 

A cross-classification table  comparison of the 6 and 7 group models is shown in Table~\ref{tab:teach:groupcomp}. This demonstrates that the main difference between the two models from a clustering perspective is the introduction of an extra group - Group 6 in the 7 group model - which contains observations clustered to Groups 1 and 3 with high uncertainty in the 6 group model. This suggests that the additional groups in the model may improve model fit without necessarily introducing additional distinct clusters to the data.

%\begin{table}[ht]
%\caption{Bayes factors for number of groups in Teaching Styles data.}
%\centering
%\begin{tabular}{rrrrrr}
%  \hline
% & $\mathcal{BF}_{6,7}$ & $\mathcal{BF}_{7,8}$ & $\mathcal{BF}_{8,9}$ & $\mathcal{BF}_{9,10}$ & $\mathcal{BF}_{10,11}$\\ 
%  \hline
% & 13.16 & 5.40 & 3.84 & 2.29 & 1.32 \\ 
%      \hline
%\end{tabular}
%\label{tab:teach:group}
%\end{table}

\begin{table}[ht]
\caption{Posterior probability for number of groups in Teaching Styles data.}
\centering
\begin{tabular}{rrrrrr}
  \hline
  $p_6$ & $p_7$ & $p_8$ & $p_9$ & $p_{10}$ & $p_{11}$\\ 
  \hline
  0.2071 & 0.3894 & 0.2627 & 0.1120 & 0.0257 & 0.0031 \\ 
      \hline
\end{tabular}
\label{tab:teach:group}
\end{table}

\begin{table*}[ht]
\centering
\caption{Comparison of mapped clusterings for the 6 and 7 group models fitted to the Teaching Styles dataset. For convenience some rows in the table have been re-arranged to illustrate the agreement in the clusterings.}
\begin{tabular}{rrrrrrr}
  \hline
 & Group 1 & Group 2 & Group 3 & Group 4 & Group 5 & Group 6 \\ 
  \hline
Group 1 &  40 &   0 &   3 &   0 &   0 &   0 \\ 
  Group 2 &   0 &  59 &   0 &   0 &   0 &   0 \\ 
  Group 3 &   0 &   0 & 105 &   0 &   0 &   0 \\ 
  Group 4 &   1 &   0 &   6 & 141 &   0 &   0 \\ 
  Group 5 &   0 &   0 &   3 &   0 &  51 &   0 \\ 
  Group 6 &   6 &   0 &   4 &   0 &   0 &   0 \\ 
  Group 7 &   0 &   0 &   1 &   0 &   0 &  47 \\ 
   \hline
\end{tabular}
\label{tab:teach:groupcomp}
\end{table*}

\begin{figure*}[t]
\begin{center}
\centering
% \subfigure{}
\subfigure{\includegraphics[width=0.47\textwidth]{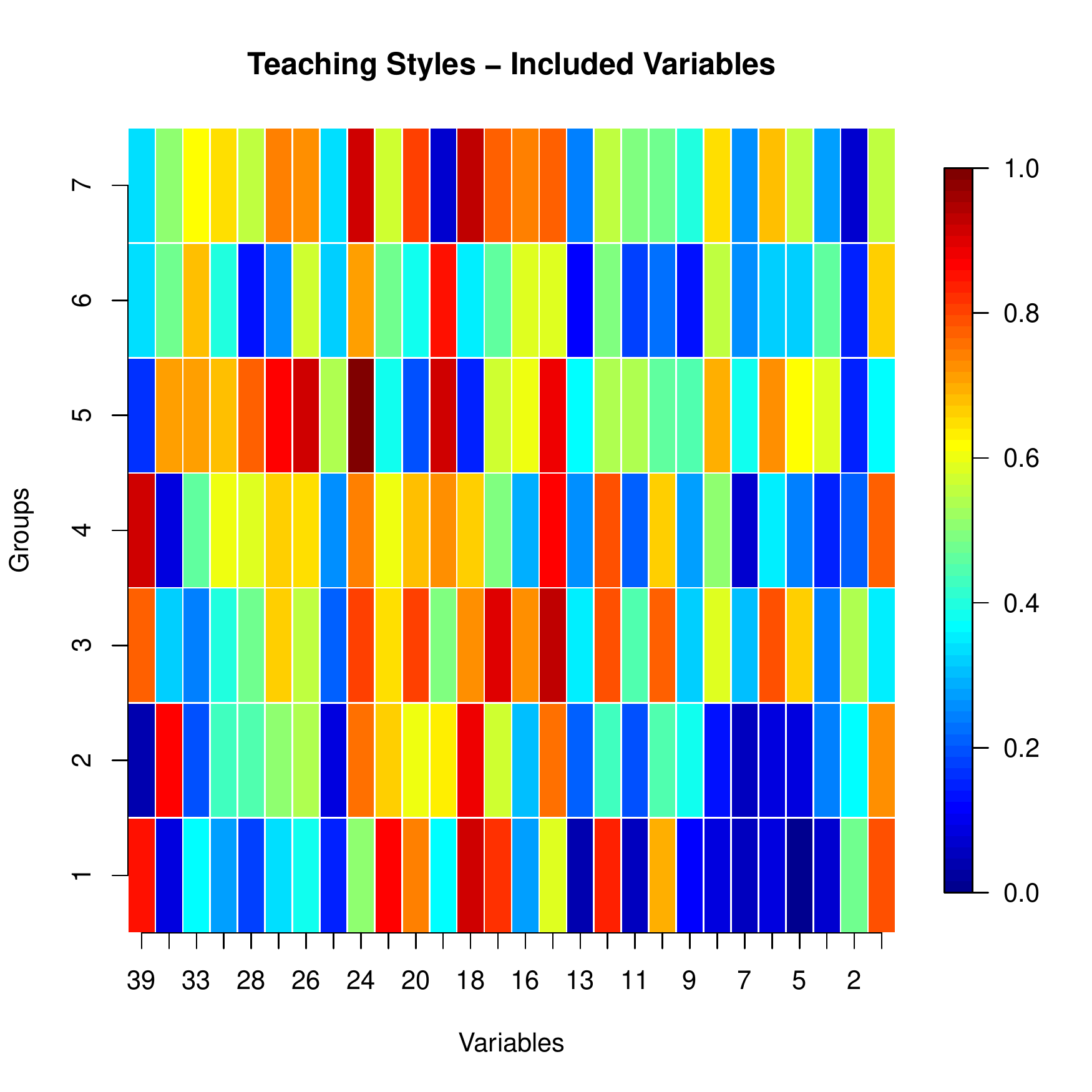}}
\subfigure{\includegraphics[width=0.47\textwidth]{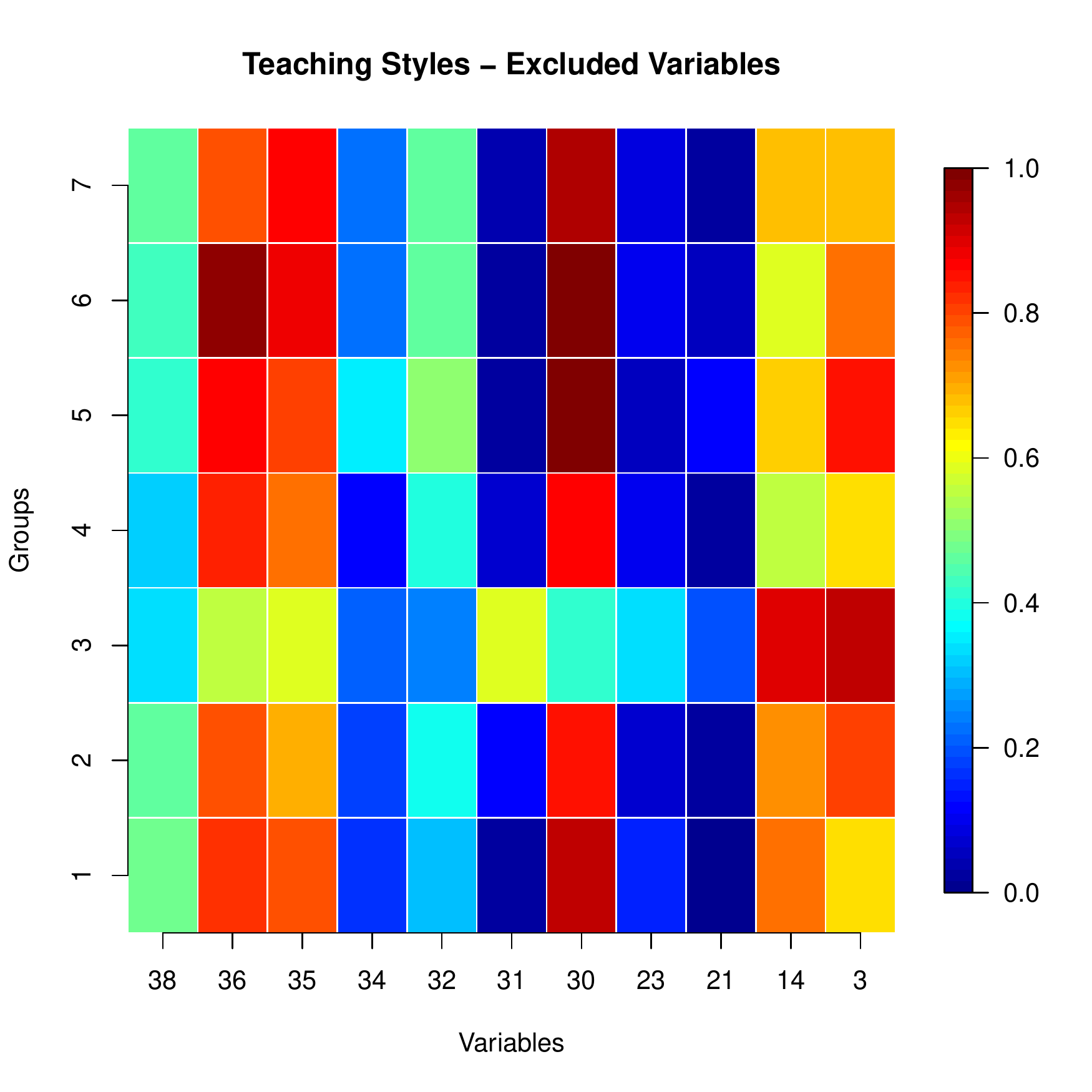}}
\end{center}
\caption{Heatmap plots of the item probability parameters for 7 group model fitted to the Teaching Styles data, for the included and excluded variables. Parameter estimates are highly similar for the excluded variables, while varied behaviour is apparent for included variables. }
\label{fig:Teach_vars}
\end{figure*}

Figure~\ref{fig:Teach_vars} compares the dropped variables to those retained by the sampler. This shows heatmap plots of the item probability parameters for the 7 group model, firstly for the included and then the excluded variables. (Note that due to space restrictions, only every second variable index is included for the left plot.) Parameter estimates are highly similar for the excluded variables, while the behaviour of the parameters is much more varied for the included variables. While the estimates of parameters in Group 3 appear different from other groups for the excluded parameters, this is the smallest group in the clustering, and as a result standard error estimates for this group's item probability parameters are extremely high, ranging between 35\% and 43\%.

\subsection{Physiotherapy Data}\label{sec:apply:physio}
\begin{figure*}[t]
\begin{center}
\centering
\subfigure{ \includegraphics[width=0.32\textwidth]{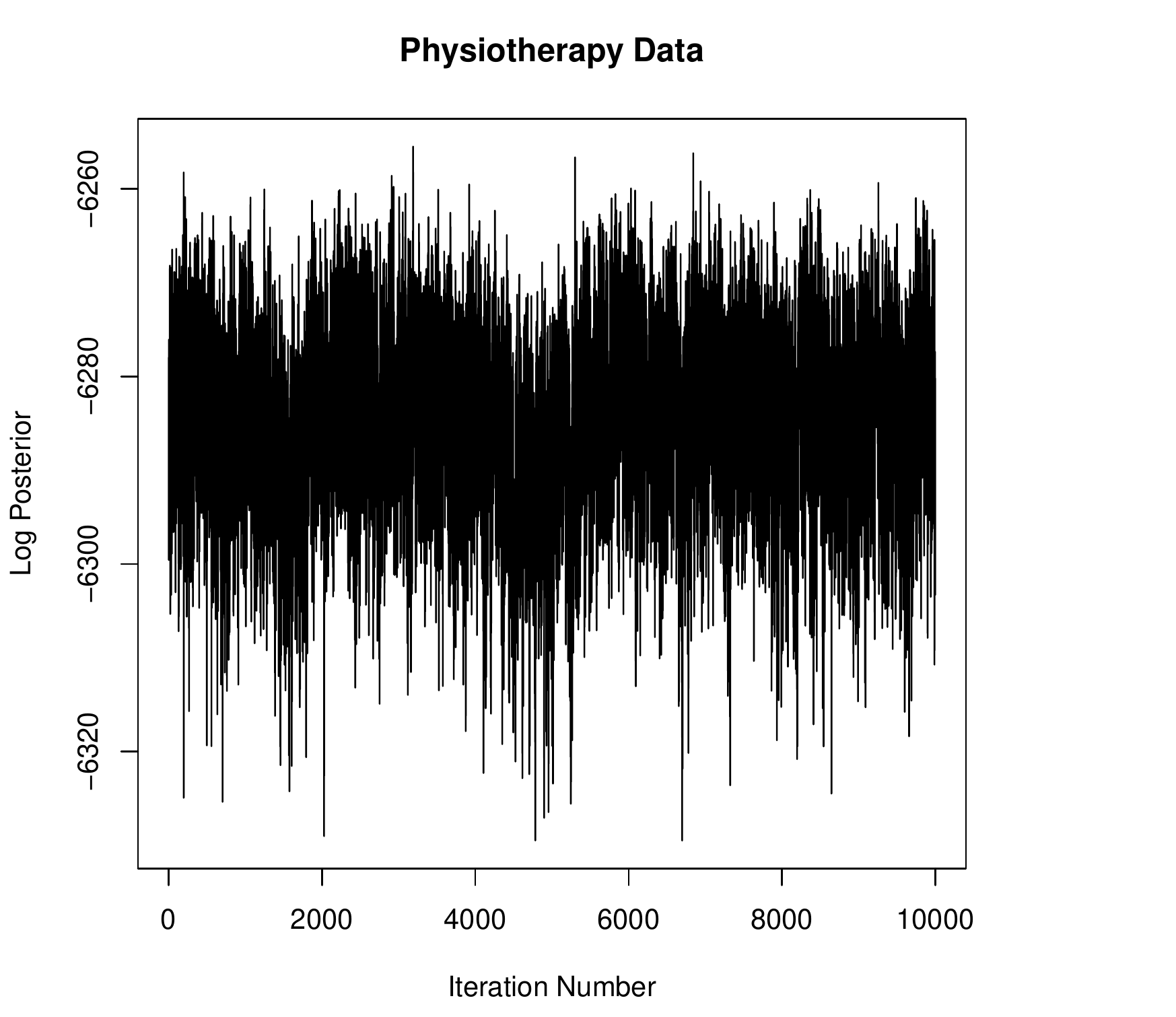}}
\subfigure{ \includegraphics[width=0.32\textwidth]{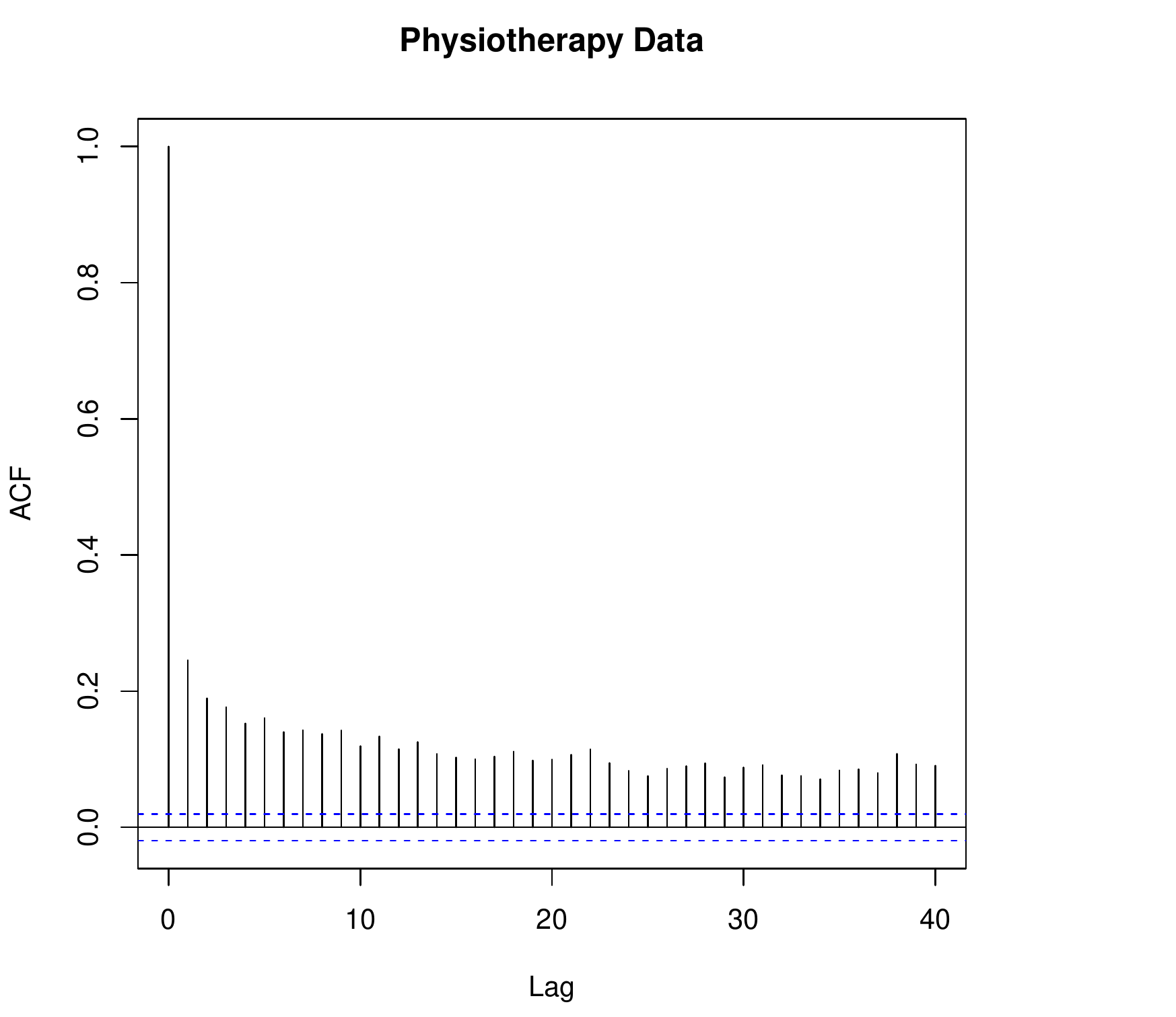}}
\subfigure{  \includegraphics[width=0.32\textwidth]{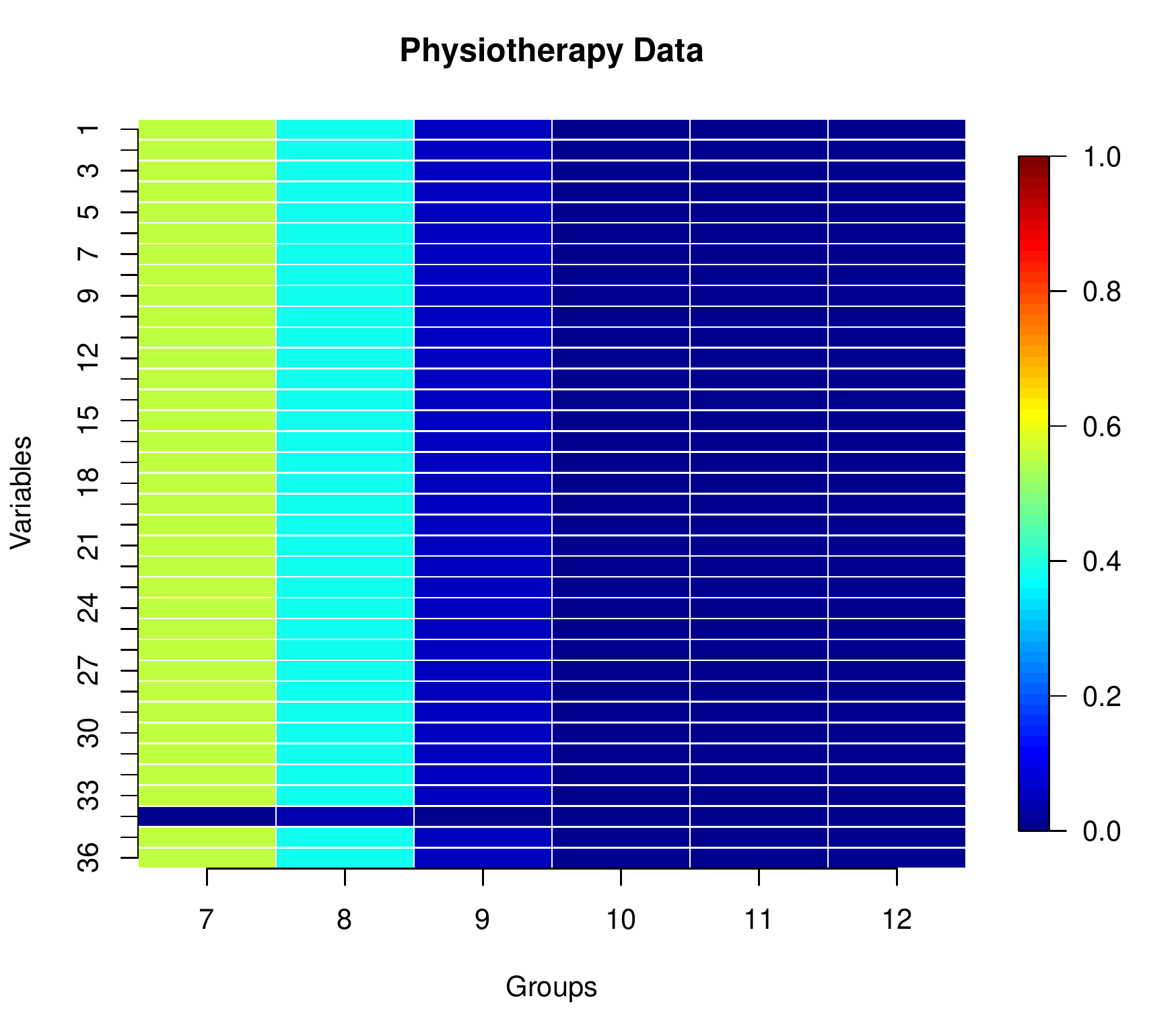}}
\end{center}
\caption{Diagnostic plots of the collapsed sampler applied to the Physiotherapy dataset, and on the right, a coincidence matrix for the sampler. }
% While the sampler doesn't appear to mix as well as in in the previous examples, it still explores the parameter space.}
\label{fig:Physio_Eval}
\end{figure*}

A collapsed sampler was run on the Physiotherapy data for 20,000 iterations after 5,000 iterations burn-in, with every second iteration retained. Auto correlation of the sampler was deemed satisfactory after these measures were taken. Diagnostic plots of the sampler are shown in Figure~\ref{fig:Physio_Eval}.

When applied to the full dataset, the sampler settles on 7 groups. (While a certain amount of evidence exists for an 8 group model, when compared, the two groupings have 98\% agreement.) This is shown in Table~\ref{tab:physio:group}. These 7 groups correspond well with the three pain types. A confusion matrix comparing the grouped observations with their reported pain types is shown in Table~\ref{tab:physio:type}. Overall there is high agreement between the two sets of groupings, with a Rand index of about 92\%. Note that some groups have extremely small mapped membership levels. Almost all variables proved useful in this clustering, with only one variable being dropped by the sampler. This is perhaps unsurprising, since the variables were identified for the study by expert recommendation.

%\begin{table}[ht]
%\caption{Bayes factors for number of groups in Physiotherapy data.}
%\centering
%\begin{tabular}{rrrrr}
%  \hline
% & $\mathcal{BF}_{6,7}$ & $\mathcal{BF}_{7,8}$ & $\mathcal{BF}_{8,9}$ & $\mathcal{BF}_{9,10}$ \\ 
%  \hline
% &613.66 &  5.14 &  0.97 & 0.72 \\ 
%   \hline
%\end{tabular}
%\label{tab:physio:group}
%\end{table}

\begin{table}[ht]
\caption{Posterior probability for number of groups in Physiotherapy data.}
\centering
\begin{tabular}{rrrrrrr}
  \hline
 & $p_7$ & $p_8$ & $p_9$ & $p_{10}$ & $p_{11}$ & $p_{12}$ \\ 
  \hline
  & 0.5577 & 0.3879 & 0.0491 & 0.0046 & 0.0006 & 0.0001 \\ 
   \hline
\end{tabular}
\label{tab:physio:group}
\end{table}

%\begin{table}[ht]
%\caption{Posterior probability for number of groups in Physiotherapy data.}
%\centering
%\begin{tabular}{rrrrr}
%  \hline
%  $p_6$ & $p_7$ & $p_8$ & $p_9$ & $p_{10}$ \\ 
%  \hline
%0.0066 & 0.5786 & 0.3718 & 0.0401 & 0.0029  \\ 
%   \hline
%\end{tabular}
%\label{tab:physio:group}
%\end{table}

\begin{table}[ht]
\centering
\caption{Confusion matrix comparing clustered observations to reported pain types in the Physiotherapy data. (CN = Central Neuropathic, N=Nociceptive, and PN = Peripheral Neuropathic.) For convenience some rows in the table have been re-arranged to illustrate the agreement in the clusterings.}
Pain Type \\~
\begin{tabular}{rrrr}
  \hline
 & CN & N & PN \\ 
  \hline
Group 1 &   3 &   0 &   1 \\ 
Group  5 &  52 &   0 &   0 \\ 
Group  7 &  30 &   3 &   0 \\ 
Group  3 &   6 &  96 &   1 \\ 
Group  6 &   0 & 120 &   1 \\ 
Group  2 &   1 &  16 &  79 \\ 
Group  4 &   3 &   0 &  13 \\ 
   \hline
\end{tabular}
\label{tab:physio:type}
\end{table}

A further study was then made to see which variables proved useful for clustering within each pain type. Samplers of length 75,000, 20,000 and 75,000 were run after 5,000 iterations of burn-in for observations reporting central neuropathic, nociceptive and peripheral neuropathic pain types respectively. For the samplers run on the  central neuropathic and peripheral neuropathic datasets, the chains were thinned in a highly conservative manner, with 1 in 15 samples retained by successive subsampling. Every second iteration was retained for the nociceptive data. 

In all cases, a high number of variables were decisively dropped by the samplers: 9, 15, and 18 variables were dropped for the respective datasets, with strong evidence that 21, 17 and 12 of the variables were informative to the clustering. No variables were excluded by all three models. The samplers identify 5, 4 and 3 group models as suitable for the data subsets, although again, several of the groups are quite small, particularly for the model fitted to the nociceptive data. 

Comparing the clusterings of the models fitted to the data subsets to the clustering of the relevant observations from the model fitted to the full dataset, shows a high level of agreement, with Rand indices of 82\%, 90\% and 93\% respectively. In particular, it is worth noting that when analyzing only the central neuropathic and nociceptive subsets, the number of groups chosen is the same as the number of groups used for observations on the respective pain type when analyzing the full dataset.  

For the peripheral neuropathic subset, only two observations with this pain type are not placed into one of three groups, the number of groups chosen by the model fitted to the data subset. 

The main difference between the clusterings is that the groups in models fitted to the data subsets are of a more equal size in comparison with the relevant subsets obtained from clustering the full dataset; this makes sense, as excluding the variables which are not meaningful for the subset in question make it easier to distinguish between different types of behaviour. A comparison of the clusterings for the nociceptive dataset is shown in Table~\ref{tab:physio:noc}.

\begin{table}[ht]
\centering
\caption{Confusion matrix comparing clustered observations based on the model fitted to the nociceptive subset to the clustering of the relevant observations obtained from the 7 group model applied to the full Physiotherapy dataset.}
 Full dataset\\Nociceptive subset
\begin{tabular}{rrrrr}
  \hline
 & Group 6 &Group 3 &Group 2 &Group 7 \\ 
  \hline
Group 2  &  107 &  2 &  0 &   0 \\ 
Group 4  &   0 &   81 & 0 &  1 \\ 
Group 1  &   13 &  4 &  15 & 0 \\ 
Group 3  &   0 &  9 &   1 &  2 \\ 
   \hline
\end{tabular}
\label{tab:physio:noc}
\end{table}

\section{Discussion}\label{sec:discuss}
In this paper we have introduced a model to perform LCA, including group and variable selection, in a Bayesian setting. Directly sampling the number of groups and the inclusion of variables allows for a suitable model to be chosen in a highly principled manner. While estimates of the posterior distribution of model parameters are not directly available, as they would be using a full model sampler, posterior means and standard deviations are nevertheless straightforward to calculate. Furthermore, in a Bayesian setting, the probabilistically driven search of the collapsed sampler over the discrete model space is a more computationally efficient approach than exhaustively computing information criteria. 

Marginalisation of the posterior can lead to a more computationally efficient algorithm, especially when clustering and model selection are the main aims of analyst. Firstly, the fact that parameter samples are not required reduces the computational burden. Secondly, the omission of parameter sampling can lead to more efficient mixing, particularly for transdimensional moves, which in the case of variable selection, may require jumping between spaces of large dimension. We investigate this further in Appendix~\ref{append:A}.

In the latter two applications described in Section~\ref{sec:apply}, some groups in the selected model contained only a small number of observations. These observations might potentially be viewed as outliers, or as being poorly described by the larger groups found by the model, rather than as distinct clusters. This is a particularly pragmatic approach to take when clustering in a model based setting, where the resultant high posterior standard deviations of model parameters for these small groups make interpretation of the group behaviour essentially meaningless. 

We note that the outlined model for variable selection is somewhat limited by the conditional independence assumption of the LCA model. In other words, when a variable is proposed to be included or excluded by the sampling scheme, we are asking the question ``does the proposed variable contain information about the clustering?'', rather than ``does the proposed variable contain \emph{additional} information about the clustering?'' Thus, while non-informative variables are removed satisfactorily, for example, two informative but highly correlated variables would both be included with high probability, when perhaps the inclusion of just one would provide clustering results of similar quality. \citet{dean06} propose such a model for variable selection when clustering continuous data; the covariance parameter of a multivariate normal distribution being a natural way to model the conditional dependence between variables for this type of data.

Latent trait analysis models \citep[see][for example]{bartholomew99} allow dependency between multivariate categorical data, and \citet{gollini2013} have recently proposed a mixture of latent trait analyzers model which does not make the conditional independence assumption of LCA. Potentially this model could prove useful for variable selection in a clustering setting. On the other hand, implementing the model for this purpose may be more difficult than for LCA, as the likelihood is not available in closed form.

Finally, in Section~\ref{sec:apply:physio} we applied the collapsed sampler to subsets of the dataset, based on additionally labelled information. In a general clustering setting, it may be of interest to determine which variables are most helpful for identifying particular groups in the data. This is a sort of unsupervised discriminant analysis which may be of future interest to further reduce the dimensionality of a dataset.

\bibliography{draft.bib}

\appendix
\section{Comparison with Reversible Jump MCMC}\label{append:A}
In this section we investigate how the performance of a collapsed Gibbs sampler compares with an RJMCMC based on using all model parameters. We divide this investigation into two tasks: selecting a) the number of classes, and b) which variables to include. We implement the approach of  \cite{JiaChiun2013} using already available software to investigate the efficacy of RJMCMC for the former task, and outline our own approach to perform the latter for the case where the observed data is binary only. We find that the approach performs reasonably well when selecting the number of classes, although its performance is somewhat slower than that of the collapsed sampler. The implemented approach performs poorly when performing variable selection.

 \subsection{ Number of Classes}\label{sec:appendgroup}
To identify the number of groups in a dataset using RJMCMC methods, we apply software\footnote{\texttt{http://ghuang.stat.nctu.edu.tw/software/download.htm}} implementing the approach of  \cite{JiaChiun2013}.  We applied the software to the binary and non-binary Dean and Raftery datasets described in Section~\ref{sec:data_description}, running the sampler for 100,000 iterations in both cases. All prior settings were set by default. In both cases, the non-informative variables were removed, since the sole task was to identify the correct number of classes.

As the software for this approach was implemented as a \textbf{\texttt{C++}} programme, it can be thought of as broadly to our own collapsed sampler, which is implemented in \textbf{\texttt{C}}; for the binary and non-binary datasets, the software took roughly 25 and 90 minutes to run respectively, based on the same hardware specifications described previously. In both cases, this was markedly longer than the collapsed sampler took, despite the fact that the model was exploring the group space only, and the dimension of the data had been reduced.

The results from the samplers are shown in Table~\ref{tab:Huang}. In the case of the binary data, the correct number of groups is chosen as the most likely candidate, although, with a lower posterior probability in comparison to the collapsed sampler. In the case of the non-binary data, $G = 2$ is incorrectly is chosen as the most likely candidate, with some level of uncertainty surrounding which model is the most suitable.

\begin{table}[ht]
\centering
\caption{Posterior probability for number of groups in the binary and non-binary Dean and Raftery datasets using the reversible jump approach of \citep{JiaChiun2013}. }
\begin{tabular}{rrrrrrr}
  \hline
 & $p_1$ & $p_2$ & $p_3$ & $p_4$ & $p_5$ & $p_{\geq 6}$\\ 
  \hline
Binary & 0.12 & 0.39 & 0.27 & 0.22 & 0 & 0  \\ 
  \hline
Non-binary & 0.09 & 0.23 & 0.14 & 0.15 & 0.13 & 0.26 \\
   \hline
\end{tabular}
\label{tab:Huang}
\end{table}

%    \begin{table}[ht]
%
%\caption{Posterior probability for number of groups in the non-binary Dean and Raftery data using the reversible jump approach of \citep{JiaChiun2013}. }
%\begin{tabular}{rrrrrrrr}
%  \hline
%   & $p_1$ & $p_2$ & $p_3$ & $p_4$ \\
%  \hline
% & 0.0924 & 0.2294 & 0.1414 & 0.1476 & 0.1314 & 0.2578 \\
%   \hline
%\end{tabular}
%%\flushleft
%%\begin{tabular}{rr}
%%  \hline
%%  
%%  \hline
%%    \\ 
%%   \hline
%%\end{tabular}
%\label{tab:Huang2}
%\end{table}     

 \subsection{Variable Selection}\label{sec:appendvariableselect}
 Recall that for the variable inclusion/exclusion step, a variable ${m^*}$ is selected at random from $1, \ldots, M.$ An inclusion or exclusion move is then proposed, based on the current status of the variable. In what follows, we assume that the state space has $G$ groups, and that the data is binary, so that $X_{nm} \in \{0, 1\}$, for all $n = 1, \ldots, N$ and $m = 1, \ldots, M.$ 
 
 \subsubsection*{ Inclusion Step}
 Suppose we select a variable ${m^*},$ which is currently excluded from the model. For the inclusion step, dropping the variable index, we propose the following move:

 \begin{enumerate}
\item Generate $u_1, \dots, u_{G-1} \sim \mbox{Uniform}(-\epsilon, \epsilon), $ and set \\ $u_{G} = - \sum^{G-1}_{i=1} u_i.$
\item Set $$ \log\left(\frac{\theta_{1}}{1 - \theta_{1}}\right) = \log\left(\frac{\beta}{1 - \beta}\right) + u_1, $$ which is equivalent to setting 
\[ \theta_1 = \frac{\beta e^{u_1}}{1 + \beta (e^{u_1} -1 )}.\]
Similarly, for $g = 2, \ldots, G,$ set: \[\theta_g = \frac{\beta e^{u_g}}{1 + \beta (e^{u_g} -1 )}.\]
\end{enumerate}
Then the proposed move is accepted with probability $\alpha$, where 

\begin{eqnarray*} \alpha &=& \min \left ( 1, \frac{  p_{\mathrm{cl}}(\bX, \bZ|\tilde{\boldtheta},\tilde{\boldnu},\boldtau,G) }{ p_{\mathrm{cl}}(\bX, \bZ|\boldtheta,\boldnu,\boldtau,G) }   \frac{p_{\mathrm{n}}(\bX|\tilde{\boldrho},\tilde{\boldnu})}{p_{\mathrm{n}}(\bX|{\boldrho},{\boldnu})}    \right .   \\ 
&\times & \frac{ \prod_{m \in \tilde{\boldnu}_{\mathrm{n}}} p(\tilde{\boldrho}_m | \beta) }{  \prod_{m \in \boldnu_{\mathrm{n}}} p(\boldrho_m | \beta)  }  \frac{ \prod_{g=1}^G \prod_{m \in \tilde{\boldnu}_{\mathrm{cl}}} p(\tilde{\boldtheta}_{gm} | \beta)  }{ \prod_{g=1}^G \prod_{m \in \boldnu_{\mathrm{cl}}} p(\boldtheta_{gm} | \beta) } \\
&\times&  \left .   \frac{ \prod_{g=1}^G \prod_{m \in \tilde{\boldnu}_{\mathrm{cl}}} p(\tilde{\boldtheta}_{gm} | \beta)  }{ \prod_{g=1}^G \prod_{m \in \boldnu_{\mathrm{cl}}} p(\boldtheta_{gm} | \beta) }  |{\mathcal J}| \times p(\xi \rightarrow \xi^*) \right),
%\frac{  p_{\mathrm{cl}}(\bX, \bZ|\tilde{\boldtheta},\tilde{\boldnu},\boldtau,G) p_{\mathrm{n}}(\bX|\tilde{\boldrho},\tilde{\boldnu}) \prod_{m \in \tilde{\boldnu}_{\mathrm{n}}} p(\tilde{\boldrho}_m | \beta)
 %\prod_{g=1}^G \prod_{m \in \tilde{\boldnu}_{\mathrm{cl}}} p(\tilde{\boldtheta}_{gm} | \beta) } 
 % {  p_{\mathrm{cl}}(\bX, \bZ|\boldtheta,\boldnu,\boldtau,G) p_{\mathrm{n}}(\bX|\boldrho,\boldnu)   \prod_{m \in \boldnu_{\mathrm{n}}} p(\boldrho_m | \beta)  \prod_{g=1}^G \prod_{m \in \boldnu_{\mathrm{cl}}} p(\boldtheta_{gm} | \beta) } \times |{\mathcal J}| \times p(\xi \rightarrow \xi^*) \right),  $$
 \end{eqnarray*}
 where 
 \begin{eqnarray*}
  \frac{  p_{\mathrm{cl}}(\bX, \bZ|\tilde{\boldtheta},\tilde{\boldnu},\boldtau,G) }{ p_{\mathrm{cl}}(\bX, \bZ|\boldtheta,\boldnu,\boldtau,G) } &=& \prod^G_{g=1} \theta_{gm}^{S_{g{m^*}}}(1 - \theta_{g{m^*}})^{S^C_{g{m^*}}}, \\
   \frac{p_{\mathrm{n}}(\bX|\tilde{\boldrho},\tilde{\boldnu})}{p_{\mathrm{n}}(\bX|{\boldrho},{\boldnu})}  &=& \frac{1}{\rho_{m^*}^{N_{m^*}}(1 - \rho_{m^*})^{N - N_{m^*}}},\\
    \frac{ \prod_{m \in \tilde{\boldnu}_{\mathrm{n}}} p(\tilde{\boldrho}_m | \beta) }{  \prod_{m \in \boldnu_{\mathrm{n}}} p(\boldrho_m | \beta)  } &=& \frac{\Gamma(\beta)^2}{\Gamma({2\beta})} \times \frac{1}{ \rho_{m^*}^{\beta -1}(1 - \rho_{m^*}^{\beta -1})},\\
    \frac{ \prod_{g=1}^G \prod_{m \in \tilde{\boldnu}_{\mathrm{cl}}} p(\tilde{\boldtheta}_{gm} | \beta)  }{ \prod_{g=1}^G \prod_{m \in \boldnu_{\mathrm{cl}}} p(\boldtheta_{gm} | \beta) } &=& G \frac{\Gamma(2 \beta)}{\Gamma(\beta)^2} \prod^G_{g=1} \theta_{gm^*}^{\beta -1} (1 - \theta_{gm^*})^{\beta -1},\\
\end{eqnarray*}
and we define  $S_{g{m^*}} = \sum^N_{n=1} X_{nm^*}Z_{ng}, $ $S^C_{g{m^*}} = \sum^N_{n=1} (1 -  X_{nm^*})Z_{ng}, $ and $N_{m^*} = \sum^N_{n=1} X_{nm^*}.$ Here we use $p(\xi \rightarrow \xi^*) = 1/M$ to denote the probability of the proposed move. Finally, the Jacobian ${\mathcal J}$ is defined as ${\mathcal J}_{1g} = \frac{\partial \theta_{gm^*}}{\partial \rho_{m^*}},$ and ${\mathcal J}_{kg} =  \frac{\partial \theta_{gm^*}}{\partial u_{k-1}},$ for $g = 1, \dots, G \mbox{ and } k = 2, \dots, G.$

\subsubsection*{ Exclusion Step}
If the variable ${m^*},$ is currently included in the model, we propose the exclusion step, \[\rho = \frac{\left (\prod^G_{g=1}  \theta_g \right )^{1/G} }{ \left ( \prod^G_{g=1}  \theta_g \right )^{1/G} + \left (\prod^G_{g=1}  (1 -  \theta_g) \right )^{1/G} },\] where again we have dropped the variable index.
Using this expression, we then obtain \[ u_g = \left( 1 - \frac{1}{G} \right ) \log \left ( \frac{ \theta_g }{ 1 - \theta_g } \right ) - \frac{1}{G} \sum_{j \neq g}   \log \left ( \frac{ \theta_j }{ 1 - \theta_j } \right ), \] for $g = 1, \ldots, G-1$, demonstrating the required bijection between $\boldtheta_{m^*}$ and $(\rho_{m^*}, \mathbf{u})$. The proposed move is again accepted with probability $\alpha$, where \begin{eqnarray*} \alpha &=& \min \left ( 1,  \frac{  p_{\mathrm{cl}}(\bX, \bZ|\tilde{\boldtheta},\tilde{\boldnu},\boldtau,G) }{ p_{\mathrm{cl}}(\bX, \bZ|\boldtheta,\boldnu,\boldtau,G) } \frac{p_{\mathrm{n}}(\bX|\tilde{\boldrho},\tilde{\boldnu})}{p_{\mathrm{n}}(\bX|{\boldrho},{\boldnu})}  \right. \\
&\times& \frac{ \prod_{m \in \tilde{\boldnu}_{\mathrm{n}}} p(\tilde{\boldrho}_m | \beta) }{  \prod_{m \in \boldnu_{\mathrm{n}}} p(\boldrho_m | \beta)  }   \frac{ \prod_{g=1}^G \prod_{m \in \tilde{\boldnu}_{\mathrm{cl}}} p(\tilde{\boldtheta}_{gm} | \beta)  }{ \prod_{g=1}^G \prod_{m \in \boldnu_{\mathrm{cl}}} p(\boldtheta_{gm} | \beta) } \\
&\times&  \left .  \frac{ \Gamma(\beta)^2 }{ G\Gamma(2 \beta) } \times \frac{1 } { \prod^G_{g=1} \theta_{gm^*}^{\beta -1} (1 - \theta_{gm^*})^{\beta -1} } |{\mathcal J}| \times p(\xi \rightarrow \xi^*) \right)
%\frac{  p_{\mathrm{cl}}(\bX, \bZ|\tilde{\boldtheta},\tilde{\boldnu},\boldtau,G) p_{\mathrm{n}}(\bX|\tilde{\boldrho},\tilde{\boldnu}) \prod_{m \in \tilde{\boldnu}_{\mathrm{n}}} p(\tilde{\boldrho}_m | \beta)
% \prod_{g=1}^G \prod_{m \in \tilde{\boldnu}_{\mathrm{cl}}} p(\tilde{\boldtheta}_{gm} | \beta) } {  p_{\mathrm{cl}}(\bX, \bZ|\boldtheta,\boldnu,\boldtau,G) p_{\mathrm{n}}(\bX|\boldrho,\boldnu)   \prod_{m \in \boldnu_{\mathrm{n}}} p(\boldrho_m | \beta)  \prod_{g=1}^G \prod_{m \in \boldnu_{\mathrm{cl}}} p(\boldtheta_{gm} | \beta) } \times |{\mathcal J}| \times p(\xi \rightarrow \xi^*) \right), 
  \end{eqnarray*}
 where the calculations are inverted, so that
 
 \begin{eqnarray*}
  \frac{  p_{\mathrm{cl}}(\bX, \bZ|\tilde{\boldtheta},\tilde{\boldnu},\boldtau,G) }{ p_{\mathrm{cl}}(\bX, \bZ|\boldtheta,\boldnu,\boldtau,G) } &=& \frac{1}{ \prod^G_{g=1} \theta_{gm}^{S_{g{m^*}}}(1 - \theta_{g{m^*}})^{S^C_{g{m^*}}} }, \\
   \frac{p_{\mathrm{n}}(\bX|\tilde{\boldrho},\tilde{\boldnu})}{p_{\mathrm{n}}(\bX|{\boldrho},{\boldnu})}  &=& \rho_{m^*}^{N_{m^*}} (1 - \rho_{m^*})^{N - N_{m^*} },\\
    \frac{ \prod_{m \in \tilde{\boldnu}_{\mathrm{n}}} p(\tilde{\boldrho}_m | \beta) }{  \prod_{m \in \boldnu_{\mathrm{n}}} p(\boldrho_m | \beta)  } &=& \frac{ \Gamma({2\beta}) }{ \Gamma(\beta)^2 } { \rho_{m^*}^{\beta -1}(1 - \rho_{m^*}^{\beta -1})},\\
    \frac{ \prod_{g=1}^G \prod_{m \in \tilde{\boldnu}_{\mathrm{cl}}} p(\tilde{\boldtheta}_{gm} | \beta)  }{ \prod_{g=1}^G \prod_{m \in \boldnu_{\mathrm{cl}}} p(\boldtheta_{gm} | \beta) } &=&  \frac{ \Gamma(\beta)^2 }{ G\Gamma(2 \beta) } \times \frac{1 } { \prod^G_{g=1} \theta_{gm^*}^{\beta -1} (1 - \theta_{gm^*})^{\beta -1} }.\\
\end{eqnarray*} The probability of the proposed move remains, $p(\xi \rightarrow \xi^*) = 1/M$.

\subsubsection*{ Dean and Raftery Data Application}
We apply this approach to the binary Dean and Raftery dataset described previously in Section~\ref{sec:data_description}. Here, we fix the number of groups to the true value  $G = 2$, so that the model search is based on variable selection only. The sampler was run for 50,000 iterations, with $\epsilon = 1$, which resulted in an acceptance probability for the inclusion/exclusion move of $\alpha \approx 0.12.$ 

The posterior probability for variable inclusion from the sampler are shown in Table~\ref{tab:appendDRvars}. None of the informative variables are selected as frequently as for the collapsed sampler, with the model only finding weak evidence for variable 1, and failing to distinguish between the other variables. 

\begin{table}[ht]
\centering
\caption{Posterior probability for variable inclusion in the binary Dean and Raftery data using RJMCMC. }
\begin{tabular}{rr}
  \hline
 & $\mathbb P(\mbox{Inclusion})$\\ 
  \hline
Variable 1 & 0.62 \\ 
Variable  2 & 0.51 \\ 
Variable  3 & 0.53 \\ 
Variable  4 & 0.55 \\ 
Variable  5 & 0.48 \\ 
Variable  6 & 0.53 \\ 
Variable  7 & 0.51 \\ 
Variable  8 & 0.46 \\ 
Variable  9 & 0.48 \\ 
Variable  10 & 0.52 \\ 
Variable  11 & 0.52 \\ 
Variable  12 & 0.50 \\ 
Variable  13 & 0.49 \\ 
   \hline
\end{tabular}
\label{tab:appendDRvars}
\end{table}

\end{document}